\input epsf
%
%
%
\def\unredoffs{} 

%
%
%
%
\newbox\leftpage \newdimen\fullhsize \newdimen\hstitle \newdimen\hsbody
\tolerance=1000\hfuzz=2pt
\catcode`\@=11 
%
\magnification=1200\unredoffs\baselineskip=16pt plus 2pt minus 1pt
\hsbody=\hsize \hstitle=\hsize 
%
%
%
\newcount\yearltd\yearltd=\year\advance\yearltd by -1900

%
%

\def\draftmode{\message{ DRAFTMODE }\def\draftdate{{\rm preliminary draft:
\number\month/\number\day/\number\yearltd\ \ \hourmin}}%
\headline={\hfil\draftdate}\writelabels\baselineskip=20pt plus 2pt minus 2pt
 {\count255=\time\divide\count255 by 60 \xdef\hourmin{\number\count255}
  \multiply\count255 by-60\advance\count255 by\time
  \xdef\hourmin{\hourmin:\ifnum\count255<10 0\fi\the\count255}}}
\def\nolabels{\def\wrlabeL##1{}\def\eqlabeL##1{}\def\reflabeL##1{}}
\def\writelabels{\def\wrlabeL##1{\leavevmode\vadjust{\rlap{\smash%
{\line{{\escapechar=` \hfill\rlap{\sevenrm\hskip.03in\string##1}}}}}}}%
\def\eqlabeL##1{{\escapechar-1\rlap{\sevenrm\hskip.05in\string##1}}}%
\def\reflabeL##1{\noexpand\llap{\noexpand\sevenrm\string\string\string##1}}}
\nolabels
%
\global\newcount\secno \global\secno=0
\global\newcount\meqno \global\meqno=1
\def\newsec#1{\global\advance\secno by1\message{(\the\secno. #1)}
\global\subsecno=0\eqnres@t\noindent{\bf\the\secno. #1}
\writetoca{{\secsym} {#1}}\par\nobreak\medskip\nobreak}
\def\eqnres@t{\xdef\secsym{\the\secno.}\global\meqno=1\bigbreak\bigskip}
\def\sequentialequations{\def\eqnres@t{\bigbreak}}\xdef\secsym{}
\global\newcount\subsecno \global\subsecno=0
\def\subsec#1{\global\advance\subsecno by1\message{(\secsym\the\subsecno. #1)}
\ifnum\lastpenalty>9000\else\bigbreak\fi
\noindent{\it\secsym\the\subsecno. #1}\writetoca{\string\quad
{\secsym\the\subsecno.} {#1}}\par\nobreak\medskip\nobreak}
\def\appendix#1#2{\global\meqno=1\global\subsecno=0\xdef\secsym{\hbox{#1.}}
\bigbreak\bigskip\noindent{\bf Appendix #1 #2}\message{(#1 #2)}
\writetoca{Appendix {#1} {#2}}\par\nobreak\medskip\nobreak}
%
%
\def\eqnn#1{\xdef #1{(\secsym\the\meqno)}\writedef{#1\leftbracket#1}%
\global\advance\meqno by1\wrlabeL#1}
\def\eqna#1{\xdef #1##1{\hbox{$(\secsym\the\meqno##1)$}}
\writedef{#1\numbersign1\leftbracket#1{\numbersign1}}%
\global\advance\meqno by1\wrlabeL{#1$\{\}$}}
\def\eqn#1#2{\xdef #1{(\secsym\the\meqno)}\writedef{#1\leftbracket#1}%
\global\advance\meqno by1$$#2\eqno#1\eqlabeL#1$$}
%
\newskip\footskip\footskip14pt plus 1pt minus 1pt 
\def\footnotefont{\ninepoint}\def\f@t#1{\footnotefont #1\@foot}
\def\f@@t{\baselineskip\footskip\bgroup\footnotefont\aftergroup\@foot\let\next}
\setbox\strutbox=\hbox{\vrule height9.5pt depth4.5pt width0pt}
\global\newcount\ftno \global\ftno=0
\def\foot{\global\advance\ftno by1\footnote{$^{\the\ftno}$}}
%
\newwrite\ftfile
\def\footend{\def\foot{\global\advance\ftno by1\chardef\wfile=\ftfile
$^{\the\ftno}$\ifnum\ftno=1\immediate\openout\ftfile=foots.tmp\fi%
\immediate\write\ftfile{\noexpand\smallskip%
\noexpand\item{f\the\ftno:\ }\pctsign}\findarg}%
\def\footatend{\vfill\eject\immediate\closeout\ftfile{\parindent=20pt
\centerline{\bf Footnotes}\nobreak\bigskip\input foots.tmp }}}
\def\footatend{}
%
%
\global\newcount\refno \global\refno=1
\newwrite\rfile
\def\ref{[\the\refno]\nref}
\def\nref#1{\xdef#1{[\the\refno]}\writedef{#1\leftbracket#1}%
\ifnum\refno=1\immediate\openout\rfile=refs.tmp\fi
\global\advance\refno by1\chardef\wfile=\rfile\immediate
\write\rfile{\noexpand\item{#1\ }\reflabeL{#1\hskip.31in}\pctsign}\findarg}
\def\findarg#1#{\begingroup\obeylines\newlinechar=`\^^M\pass@rg}
{\obeylines\gdef\pass@rg#1{\writ@line\relax #1^^M\hbox{}^^M}%
\gdef\writ@line#1^^M{\expandafter\toks0\expandafter{\striprel@x #1}%
\edef\next{\the\toks0}\ifx\next\em@rk\let\next=\endgroup\else\ifx\next\empty%
\else\immediate\write\wfile{\the\toks0}\fi\let\next=\writ@line\fi\next\relax}}
\def\striprel@x#1{} \def\em@rk{\hbox{}}
\def\lref{\begingroup\obeylines\lr@f}
\def\lr@f#1#2{\gdef#1{\ref#1{#2}}\endgroup\unskip}
\def\semi{;\hfil\break}
\def\addref#1{\immediate\write\rfile{\noexpand\item{}#1}} 
\def\startrefs#1{\immediate\openout\rfile=refs.tmp\refno=#1}
\def\xref{\expandafter\xr@f}\def\xr@f[#1]{#1}
\def\refs#1{\count255=1[\r@fs #1{\hbox{}}]}
\def\r@fs#1{\ifx\und@fined#1\message{reflabel \string#1 is undefined.}%
\nref#1{need to supply reference \string#1.}\fi%
\vphantom{\hphantom{#1}}\edef\next{#1}\ifx\next\em@rk\def\next{}%
\else\ifx\next#1\ifodd\count255\relax\xref#1\count255=0\fi%
\else#1\count255=1\fi\let\next=\r@fs\fi\next}
%

%
\newwrite\ffile\global\newcount\figno \global\figno=1
\def\fig{fig.~\the\figno\nfig}
\def\nfig#1{\xdef#1{fig.~\the\figno}%
\writedef{#1\leftbracket fig.\noexpand~\the\figno}%
\ifnum\figno=1\immediate\openout\ffile=figs.tmp\fi\chardef\wfile=\ffile%
\immediate\write\ffile{\noexpand\medskip\noexpand\item{Fig.\ \the\figno. }
\reflabeL{#1\hskip.55in}\pctsign}\global\advance\figno by1\findarg}
\def\xfig{\expandafter\xf@g}\def\xf@g fig.\penalty\@M\ {}
\def\figs#1{figs.~\f@gs #1{\hbox{}}}
\def\f@gs#1{\edef\next{#1}\ifx\next\em@rk\def\next{}\else
\ifx\next#1\xfig #1\else#1\fi\let\next=\f@gs\fi\next}
\newwrite\lfile
{\escapechar-1\xdef\pctsign{\string\%}\xdef\leftbracket{\string\{}
\xdef\rightbracket{\string\}}\xdef\numbersign{\string\#}}

\def\writestop{\def\writestoppt{\immediate\write\lfile{\string\pageno%
\the\pageno\string\startrefs\leftbracket\the\refno\rightbracket%
\string\def\string\secsym\leftbracket\secsym\rightbracket%
\string\secno\the\secno\string\meqno\the\meqno}\immediate\closeout\lfile}}
\def\writestoppt{}\def\writedef#1{}
\def\seclab#1{\xdef #1{\the\secno}\writedef{#1\leftbracket#1}\wrlabeL{#1=#1}}
\def\subseclab#1{\xdef #1{\secsym\the\subsecno}%
\writedef{#1\leftbracket#1}\wrlabeL{#1=#1}}
\newwrite\tfile \def\writetoca#1{}
\def\leaderfill{\leaders\hbox to 1em{\hss.\hss}\hfill}
\def\writetoc{\immediate\openout\tfile=toc.tmp
   \def\writetoca##1{{\edef\next{\write\tfile{\noindent ##1
   \string\leaderfill {\noexpand\number\pageno} \par}}\next}}}

%
\catcode`\@=12 
%
\edef\tfontsize{\ifx\answ\bigans scaled\magstep3\else scaled\magstep4\fi}
\font\titlerm=cmr10 \tfontsize \font\titlerms=cmr7 \tfontsize
\font\titlermss=cmr5 \tfontsize \font\titlei=cmmi10 \tfontsize
\font\titleis=cmmi7 \tfontsize \font\titleiss=cmmi5 \tfontsize
\font\titlesy=cmsy10 \tfontsize \font\titlesys=cmsy7 \tfontsize
\font\titlesyss=cmsy5 \tfontsize \font\titleit=cmti10 \tfontsize
\skewchar\titlei='177 \skewchar\titleis='177 \skewchar\titleiss='177
\skewchar\titlesy='60 \skewchar\titlesys='60 \skewchar\titlesyss='60
\def\titlefont{\def\rm{\fam0\titlerm}
\textfont0=\titlerm \scriptfont0=\titlerms \scriptscriptfont0=\titlermss
\textfont1=\titlei \scriptfont1=\titleis \scriptscriptfont1=\titleiss
\textfont2=\titlesy \scriptfont2=\titlesys \scriptscriptfont2=\titlesyss
\textfont\itfam=\titleit \def\it{\fam\itfam\titleit}\rm}
 \ifx\answ\bigans\else scaled\magstep1\fi
\ifx\answ\bigans\else

 \font\absi=cmmi10 scaled\magstep1
\font\absis=cmmi7 scaled\magstep1 \font\absiss=cmmi5 scaled\magstep1
\font\abssy=cmsy10 scaled\magstep1 \font\abssys=cmsy7 scaled\magstep1
\font\abssyss=cmsy5 scaled\magstep1 
\skewchar\absi='177 \skewchar\absis='177 \skewchar\absiss='177
\skewchar\abssy='60 \skewchar\abssys='60 \skewchar\abssyss='60
\fi
\font\ninerm=cmr9 \font\sixrm=cmr6 \font\ninei=cmmi9 \font\sixi=cmmi6
\font\ninesy=cmsy9 \font\sixsy=cmsy6 \font\ninebf=cmbx9
\font\nineit=cmti9 \font\ninesl=cmsl9 \skewchar\ninei='177
\skewchar\sixi='177 \skewchar\ninesy='60 \skewchar\sixsy='60
\def\ninepoint{\def\rm{\fam0\ninerm}
\textfont0=\ninerm \scriptfont0=\sixrm \scriptscriptfont0=\fiverm
\textfont1=\ninei \scriptfont1=\sixi \scriptscriptfont1=\fivei
\textfont2=\ninesy \scriptfont2=\sixsy \scriptscriptfont2=\fivesy
\textfont\itfam=\ninei \def\it{\fam\itfam\nineit}\def\sl{\fam\slfam\ninesl}%
\textfont\bffam=\ninebf \def\bf{\fam\bffam\ninebf}\rm}
%
%
\def\noblackbox{\overfullrule=0pt}
\hyphenation{anom-aly anom-alies coun-ter-term coun-ter-terms}
\def\inv{^{\raise.15ex\hbox{${\scriptscriptstyle -}$}\kern-.05em 1}}

\def\Dsl{\,\raise.15ex\hbox{/}\mkern-13.5mu D} 
\def\dsl{\raise.15ex\hbox{/}\kern-.57em\partial}

\def\lspace{\ifx\answ\bigans{}\else\qquad\fi}
\def\lbspace{\ifx\answ\bigans{}\else\hskip-.2in\fi} 
\def\boxeqn#1{\vcenter{\vbox{\hrule\hbox{\vrule\kern3pt\vbox{\kern3pt
        \hbox{${\displaystyle #1}$}\kern3pt}\kern3pt\vrule}\hrule}}}
\def\mbox#1#2{\vcenter{\hrule \hbox{\vrule height#2in
                \kern#1in \vrule} \hrule}}  
%
   
 \def\CH{{\cal H}}

\def\darr#1{\raise1.5ex\hbox{$\leftrightarrow$}\mkern-16.5mu #1}

\def\half{{\textstyle{1\over2}}} 
\def\roughly#1{\raise.3ex\hbox{$#1$\kern-.75em\lower1ex\hbox{$\sim$}}}
\hyphenation{Mar-ti-nel-li}

\def\Re{\,\hbox{Re}\,}

\def\1{\;1\!\!\!\! 1\;}

\def\etal{{\it et al.}}

\def\tozero#1{\mathrel{\mathop{\sim}\limits_{\scriptscriptstyle
{#1\rightarrow0 }}}}
\def\frac#1#2{{{#1}\over {#2}}}
\def\half{\hbox{${1\over 2}$}}
\def\quarter{\hbox{${1\over 4}$}}
\def\smallfrac#1#2{\hbox{${{#1}\over {#2}}$}}
\def\lnx{\ln\smallfrac{1}{x}}

\def\MS{\hbox{$\overline{\rm MS}$}}

\catcode`@=11 
\def\slash#1{\mathord{\mathpalette\c@ncel#1}}
 \def\c@ncel#1#2{\ooalign{$\hfil#1\mkern1mu/\hfil$\crcr$#1#2$}}
\def\lsim{\mathrel{\mathpalette\@versim<}}
\def\gsim{\mathrel{\mathpalette\@versim>}}
 \def\@versim#1#2{\lower0.2ex\vbox{\baselineskip\z@skip\lineskip\z@skip
       \lineskiplimit\z@\ialign{$\m@th#1\hfil##$\crcr#2\crcr\sim\crcr}}}
\catcode`@=12 

\def\PR{{\it Phys.~Rev.~}}

\def\NP{{\it Nucl.~Phys.~}}

\def\PL{{\it Phys.~Lett.~}}

\def\SJNP{{\it Sov.~Jour.~Nucl.~Phys.~}}
\def\SPJETP{{\it Sov.~Phys.~J.E.T.P.~}}

\def\JHEP{{\it Jour.~High~Energy~Phys.~}}
\def\vol#1{{\bf #1}}\def\vyp#1#2#3{\vol{#1} (#2) #3}

\def\as{\alpha_s}
\def\ahat{\hat\as}

\def\ash{\widehat\alpha_s}

\noblackbox
\pageno=0\nopagenumbers\tolerance=10000\hfuzz=5pt
\baselineskip 12pt
\line{\hfill CERN-PH-TH/2005-174}
\line{\hfill  RM3-TH/05-15}
\line{\hfill Edinburgh 2005/22}
\line{\hfill IFUM-843/FT}
\vskip 12pt
\centerline{\titlefont Perturbatively Stable Resummed Small $x$
Evolution Kernels}
\vskip 36pt\centerline{Guido~Altarelli,$^{(a,\>b)}$
Richard D.~Ball$^{(c)}$ and Stefano Forte$^{(d)}$}
\vskip 12pt
\centerline{\it ${}^{(a)}$ CERN, Department of Physics, Theory Division}
\centerline{\it CH-1211 Gen\`eve 23, Switzerland}
\vskip 6pt
\centerline{\it ${}^{(b)}$ Dipartimento di Fisica ``E.Amaldi'', 
Universit\`a Roma Tre}
\centerline{\it INFN, Sezione di Roma Tre}
\centerline{\it via della Vasca Navale 84, I--00146 Roma, Italy}
\vskip 6pt
\centerline{\it ${}^{(c)}$School of Physics, University of Edinburgh}
\centerline{\it  Edinburgh EH9 3JZ, Scotland}
\vskip 6pt
\centerline {\it ${}^{(d)}$Dipartimento di  Fisica, Universit\`a di
Milano and}
\centerline{\it INFN, Sezione di Milano, Via Celoria 16, I-20133 Milan, Italy}
\vskip 40pt
\centerline{\bf Abstract}
{\narrower\baselineskip 10pt
\medskip\noindent 
We present a  small $x$ resummation for the GLAP anomalous dimension and
its corresponding dual BFKL kernel,
which includes all the available
perturbative information and nonperturbative
constraints.  Specifically, it includes all the information coming from
next-to-leading order GLAP anomalous dimensions and BFKL kernels, from the constraints of momentum conservation, 
from renormalization-group improvement of the running coupling and
from gluon interchange
symmetry. The ensuing evolution kernel has a uniformly stable
perturbative expansion. It is very close to the unresummed NLO GLAP
kernel in most of the HERA  kinematic region, the small $x$ BFKL behaviour
being softened by momentum conservation and the running of the
coupling. Next-to-leading corrections are small thanks to the
constraint of gluon interchange symmetry. This result subsumes all
previous resummations in that it combines optimally all the
information contained in them. 
}
\vfill
\line{CERN-PH-TH/2005-174\hfill }
\line{December 2005\hfill}
\eject \footline={\hss\tenrm\folio\hss}

\lref\brus  {R.~D.~Ball and S.~Forte,
  {\tt hep-ph/9805315.}
}
\lref\comm{ R.~D.~Ball and S.~Forte,
  {\tt hep-ph/0601049}.
}
\lref\haidt{D.~Haidt, Talk at DIS 2003, St.~Petersburg, Russia, April
2003.}
\lref\glap{
V.N.~Gribov and L.N.~Lipatov,
\SJNP\vyp{15}{1972}{438}\semi  
L.N.~Lipatov, \SJNP\vyp{20}{1975}{95}\semi    
G.~Altarelli and G.~Parisi,
\NP\vyp{B126}{1977}{298}\semi  
see also
Y.L.~Dokshitzer,
{\it Sov.~Phys.~JETP~}\vyp{46}{1977}{691}.} 
\lref\nlo{G.~Curci, W.~Furma\'nski and R.~Petronzio,
\NP\vyp{B175}{1980}{27}\semi 
E.G.~Floratos, C.~Kounnas and R.~Lacaze,
\NP\vyp{B192}{1981}{417}.} 
\lref\nnlo{S.A.~Larin, T.~van~Ritbergen, J.A.M.~Vermaseren,
\NP\vyp{B427}{1994}{41}\semi  
S.A.~Larin \etal, \NP\vyp{B492}{1997}{338}.} 
\lref\bfkl{L.N.~Lipatov,
\SJNP\vyp{23}{1976}{338}\semi 
 V.S.~Fadin, E.A.~Kuraev and L.N.~Lipatov,
\PL\vyp{60B}{1975}{50}; 
 {\it Sov. Phys. JETP~}\vyp{44}{1976}{443}; 
\vyp{45}{1977}{199}\semi 
 Y.Y.~Balitski and L.N.Lipatov,
\SJNP\vyp{28}{1978}{822}.} 
\lref\CH{
S.~Catani and F.~Hautmann,
\PL\vyp{B315}{1993}{157}; 
\NP\vyp{B427}{1994}{475}.} 
\lref\fl{V.S.~Fadin and L.N.~Lipatov,
\PL\vyp{B429}{1998}{127}.  
}
\lref\fleta{
V.S.~Fadin et al, \PL\vyp{B359}{1995}{181}; 
\PL\vyp{B387}{1996}{593}; 
\NP\vyp{B406}{1993}{259}; 
\PR\vyp{D50}{1994}{5893}; 
\PL\vyp{B389}{1996}{737};  
\NP\vyp{B477}{1996}{767};  
\PL\vyp{B415}{1997}{97};  
\PL\vyp{B422}{1998}{287}.} 
\lref\cc{G.~Camici and M.~Ciafaloni,
\PL\vyp{B412}{1997}{396}; 
\PL\vyp{B430}{1998}{349}.} 
\lref\dd{V.~del~Duca, \PR\vyp{D54}{1996}{989};
\PR\vyp{D54}{1996}{4474}\semi 
V.~del~Duca and C.R.~Schmidt,
\PR\vyp{D57}{1998}{4069}\semi 
Z.~Bern, V.~del~Duca and C.R.~Schmidt,
\PL\vyp{B445}{1998}{168}.}
\lref\ross{
D.~A.~Ross,
Phys.\ Lett.\ B {\bf 431}, 161 (1998). 
}
\lref\jar{T.~Jaroszewicz,
\PL\vyp{B116}{1982}{291}.}
\lref\ktfac{S.~Catani, F.~Fiorani and G.~Marchesini,
\PL\vyp{B336}{1990}{18}\semi 
S.~Catani et al.,
\NP\vyp{B361}{1991}{645}.}
\lref\summ{R.~D.~Ball and S.~Forte,
\PL\vyp{B351}{1995}{313}\semi  
R.K.~Ellis, F.~Hautmann and B.R.~Webber,
\PL\vyp{B348}{1995}{582}.}
\lref\afp{R.~D.~Ball and S.~Forte,
\PL\vyp{B405}{1997}{317}.}
\lref\DGPTWZ{A.~De~R\'ujula {\it et al.},
\PR\vyp{D10}{1974}{1649}.}
\lref\das{R.~D.~Ball and S.~Forte,
\PL\vyp{B335}{1994}{77}; 
\vyp{B336}{1994}{77}\semi 
{\it Acta~Phys.~Polon.~}\vyp{B26}{1995}{2097}.}
\lref\kis{
See {\it e.g.}  R.~K.~Ellis, W.~J.~Stirling and B.~R.~Webber,
``QCD and Collider Physics'' (C.U.P., Cambridge 1996).}
\lref\hone{H1 Collab., {\it Eur.\ Phys.\ J.} C {\bf 21} (2001)
33.}
\lref\zeus{ZEUS Collab., {\it Eur.\ Phys.\ J.}
 C {\bf 21} (2001) 443.} 
\lref\mom{R.~D.~Ball and S.~Forte, {\it Phys. Lett.} {\bf
B359}, 362 (1995).}
\lref\bfklfits{R.~D.~Ball and S.~Forte,
{\tt hep-ph/9607291}\semi 
I.~Bojak and M.~Ernst, \PL\vyp{B397}{1997}{296};
\NP\vyp{B508}{1997}{731}\semi
J.~Bl\"umlein  and A.~Vogt,
\PR\vyp{D58}{1998}{014020}.}
\lref\flph{R.~D.~Ball  and S.~Forte,
{\tt hep-ph/9805315}\semi 
J. Bl\"umlein et al.,
{\tt hep-ph/9806368}.}
\lref\salam{G.~Salam, \JHEP\vyp{9807}{1998}{19}.}
\lref\sxap{R.~D.~Ball and S.~Forte,
\PL\vyp{B465}{1999}{271}.}
\lref\sxres{G. Altarelli, R.~D. Ball and S. Forte,
\NP{\bf B575}, 313 (2000);  
see also {\tt hep-ph/0001157}.
}
\lref\sxphen{G. Altarelli, R.~D.~Ball and S. Forte,
\NP\vyp{B599}{2001}{383};  
see also {\tt hep-ph/0104246}.}  
\lref\ciaf{M.~Ciafaloni and D.~Colferai,
\PL\vyp{B452}{1999}{372}. 
}
\lref\ciafresa{
 M.~Ciafaloni, D.~Colferai and G.~P.~Salam,
  Phys.\ Rev.\ D {\bf 60} (1999) 114036.
}
\lref\ciafresb{M.~Ciafaloni, D.~Colferai, G.~P.~Salam and A.~M.~Stasto,
  Phys.\ Rev.\ D {\bf 66} (2002) 054014; 
  Phys.\ Lett.\ B {\bf 576} (2003) 143;
  Phys.\ Rev.\ D {\bf 68} (2003) 114003,
}

\lref\heralhc{ M.~Dittmar {\it et al.},
  {\tt hep-ph/0511119.}
}
\lref\heralhcres{G.~Altarelli {\it et al.}, ``Resummation'', in M.~Dittmar {\it et al.},
  {\tt hep-ph/0511119.}} 
\lref\ciafdip{M.~Ciafaloni, D.~Colferai, G.~P.~Salam and A.~M.~Stasto,
  Phys.\ Lett.\ B {\bf 587} (2004) 87; see also
 G.~P.~Salam,
{\tt hep-ph/0501097}.

}
\lref\Liprun{L.N.~Lipatov,
\SPJETP\vyp{63}{1986}{5}.}
\lref\ColKwie{
J.~C.~Collins and J.~Kwiecinski, \NP\vyp{B316}{1989}{307}.}
\lref\CiaMue{
M.~Ciafaloni, M.~Taiuti and A.~H.~Mueller,
{\tt hep-ph/0107009}.
}
\lref\ciafac{
M.~Ciafaloni, D.~Colferai and G.~P.~Salam,
JHEP {\bf 0007}  (2000) 054
}
\lref\ciafrun{G.~Camici and M.~Ciafaloni,
\NP\vyp{B496}{1997}{305}.}
\lref\Haak{L.~P.~A.~Haakman, O.~V.~Kancheli and
J.~H.~Koch \NP\vyp{B518}{1998}{275}.} 
\lref\Bartels{N. Armesto, J. Bartels and M.~A.~Braun,
\PL\vyp{B442}{1998}{459}.} 
\lref\Thorne{R.~S.~Thorne,
\PL\vyp{B474}{2000}{372}; {\it Phys.\ Rev.} {\bf D64} (2001) 074005.
}
\lref\anders{
J.~R.~Andersen and A.~Sabio Vera,
{\tt arXiv:hep-ph/0305236.}
}
\lref\mf{
S.~Forte and R.~D.~Ball,
AIP Conf.\ Proc.\  {\bf 602} (2001) 60
{\tt hep-ph/0109235.}
}
\lref\sxrun{
G.~Altarelli, R.~D.~Ball and S.~Forte,
Nucl.\ Phys.\ B {\bf 621} (2002)  359.
}
\lref\ciafqz{
  M.~Ciafaloni,
  Phys.\ Lett.\ B {\bf 356}, 74 (1995).
}
\lref\ciafrc{M.~Ciafaloni, M.~Taiuti and A.~H.~Mueller,
{\it Nucl.\ Phys.}  {\bf B616} (2001) 349\semi 
M.~Ciafaloni et al., {\it Phys. Rev.} {\bf D66} (2002)
054014 
}
\lref\runph{G.~Altarelli, R.~D.~Ball and S.~Forte,
Nucl.\ Phys.\ B {\bf 674} (2003) 459;
see also   
 {\tt hep-ph/0310016.}
}
\lref\sxsym{G.~Altarelli, R.~D.~Ball and S.~Forte,
  Nucl.\ Phys.\ Proc.\ Suppl.\  {\bf 135} (2004) 163
}
\lref\nnlo{ S.~Moch, J.~A.~M.~Vermaseren and A.~Vogt,
  Nucl.\ Phys.\ B {\bf 691} (2004) 129\semi
 A.~Vogt, S.~Moch and J.~A.~M.~Vermaseren,
  Nucl.\ Phys.\ B {\bf 688} (2004) 101.
}
\lref\mrst{See e.g. R.~S.~Thorne, A.~D.~Martin, R.~G.~Roberts and W.~J.~Stirling,
  {\tt hep-ph/0507015.}
}
\lref\commutator{R.~D.~Ball and S.~Forte, {\it in preparation}}
\newsec{Introduction}
\noindent 
The theory of  scaling violations of deep inelastic
structure functions at small $x$ has progressed considerably over the
last few years. This theory has been characterized by the
long-standing puzzle posed by the
fact that no clear deviation of the data from the next-to-leading
order prediction has been 
observed (for instance at HERA~\heralhc), even though higher order 
perturbative corrections
grow rapidly in the small $x$ region. The recent  determination~\nnlo\ of
the full set of three--loop (NNLO) splitting functions makes a resolution of
this puzzle less of an academic exercise. Indeed, results obtained
using these NNLO splitting functions already~\mrst\ show sign of the known
small $x$ instability, for instance in the extraction of parton
distributions. Hence, NNLO phenomenology becomes problematic unless
this instability is treated in some way. 

However, the reason why superficially large
small $x$ corrections sum up to 
very moderate effects is now essentially understood. It is  the  consequence
of two effects: first, momentum conservation which provides a
constraint which limits the unbounded growth of small $x$ kernels, and
second, renormalization group improvement which through  the running
of the coupling softens the asymptotic small $x$ growth. A systematic
theory of small $x$ resummation which includes both these effects has
been developed in
refs.~\refs{\sxap\sxres\sxphen\sxrun{--}\runph}. In particular,
we have   
constructed a relatively simple analytical expression for the
improved anomalous dimension which we argue is the best way to put
together the information from LO BFKL and NLO GLAP. 
A shortcoming of the approach developed in these references is that
even orders of the resummed perturbative expansion are subject to a
large model dependence. This is due to the fact that the constraint of
momentum conservation stabilizes the perturbative expansion of
evolution kernels to all orders, but only in the collinear
region. In the ``anticollinear'' region, where parton virtualities
have reverse ordering (and thus only contribute due to $\ln {1\over
x}$ enhancement) the instability remains. As a consequence, the
stabilization of the BFKL kernel is lost in the vicinity of its
minimum, where the collinear and anticollinear regions weigh equally:
the minimum is in fact only present at leading order of the perturbative
expansion. Because the minimum is necessary for running coupling
resummation, the inclusion of NLO BFKL resummation prevents it, unless
one makes some assumption on the restoration of the minimum by higher
order corrections; this, however, introduces a sizable
model--dependence~\runph.

A similar approach to small $x$ resummation which also embodies
the principles of momentum conservation and renormalization group
improvement has been  presented in
refs.~\refs{\salam\ciafresa{--}\ciafresb}. This approach
differs~\heralhcres\  from
ours esssentially because it has the somewhat wider goal of
determining the off-shell gluon Green function, and it is therefore
rooted in the BFKL equation. Whereas this gives for example an exact
treatment of small $x$ running coupling corrections, it only allows a
numerical extraction of anomalous dimensions through deconvolution
from a given boundary condition. Also, it does not naturally lead to a
symmetric (double) leading log expansion of the anomalous dimension
(for example, when $n_f\not = 0$ the inclusion of subleading GLAP
terms in this approach has not yet been accomplished). However, 
in this approach it
was pointed out that the underlying symmetry upon the interchange of
gluon virtualities  
in the gluon-gluon scattering process which determines the BFKL kernel
can be used for the treatment of the instability in the anticollinear
region. 
This  symmetry interchanges the collinear and anticollinear regions,
and therefore it enables the implementation of the momentum
conservation constraint even in the anticollinear region. 

Im this paper, we exploit the symmetric
collinear-anticollinear resummation~\ciafresa\ to stabilize the
resummed and renormalization-group improved perturbative expansion of
refs.~\refs{\sxap\sxres\sxphen\sxrun{--}\runph}. 
This can be done by  using the
duality relations~\refs{\sxap{,}\sxres} which connect GLAP and BFKL
kernels in order to translate the symmetry constraints on BFKL
evolution into constraints on subleading contributions to the GLAP
anomalous dimensions. 
Because the ``double leading''
perturbative expansion of refs.~\sxres\ systematically includes
leading, next-to-leading,\dots logs of $Q^2$ and $\smallfrac{1}{x}$
simultaneously, its symmetrization must be done by inclusion of terms
which are subleading order by order, and 
 this requires a subtle
treatment of double counting terms. Also, implementation of the
symmetry is nontrivial when the coupling runs, because the choice of
running coupling scale may break it. Finally, 
after symmetrization, even at LO, the BFKL intercept,
which determines the asymptotic small $x$ behaviour, depends
nonlinearly on the strong coupling, and this makes
the
renormalization-group improvement of refs.~\refs{\sxrun{,}\runph}
rather more elaborate: it is still possible to derive an
asymptotically exact solution to the running-coupling BFKL equation in
terms of special functions,
but this now requires a double expansion about the BFKL minimum.
Once this is done, a stable perturbative expansion is obtained, and in
particular the NLO resummed result turns out~\refs{\sxsym{,}\ciafdip}
to be both close to the 
LO resummed one, and also quite close to the NLO fixed--order result,
thereby explaining the lack of experimental evidence for large
deviations from low fixed--order predictions at small $x$: in fact,
the resummed NLO splitting function
is closer to the NLO than to the NNLO fixed order one.

In this paper we discuss the resummation of anomalous dimensions and
splitting functions.
To make direct connection to the structure functions we need to also
implement  coefficient functions and their resummation. This also
entails a scheme choice: here, we adopt the so-called $Q_0$
scheme~\ciafqz, which coincides with the standard \MS\ scheme at fixed
order but differs from it at the resummed level. Issues of scheme
dependence were discussed in detail in ref.~\sxphen, where resummed
coefficient functions and fits to the data were explicitly
presented. The relation of $Q_0$ and \MS\ after small $x$ running
coupling resummation was discussed in ref.~\sxrun. Whereas the use of the
results presented in this paper for actual fits to data will require the
formalism presented in \refs\sxphen,\sxrun, all the important small $x$
resummation effects are included in the $Q_0$-scheme results 
presented here. In order to minimize issues of
scheme dependence, in this paper we will determine all anomalous
dimensions and splitting functions with $n_f=0$.

The paper is organized as follows. After recalling in  section~2 some
background results from the resummation formalism of
refs.~\refs{\sxap\sxres\sxphen\sxrun{--}\runph}, which will be needed
in the sequel, in section~3 we explain how the underlying
symmetry of the BFKL kernel can be coupled to the double--leading
expansion of ref.~\sxap\ in order to produce  a stable resummation at
the fixed coupling level. The formalism which is required to
generalize  the running coupling resummation
of ref.~\sxrun\ to this symmetrized case is introduced in section 4. Resummed
anomalous dimensions and splitting functions are explicitly
constructed in section~5 at the leading order and in section~6 at the
next-to-leading order.  Results for anomalous dimensions and
splitting functions are presented, discussed and compared in section~7.

\newsec{Background}

The behaviour of structure functions at small $x$ is dominated by the large
eigenvalue of evolution in the  singlet sector.
Thus we consider the singlet parton density $G(\xi,t)$
with $\xi=\log{1/x}$,
$t=\log{Q^2/\mu^2}$,
such that, for each
moment
\eqn\Nmom{  G(N,t)=\int^{\infty}_{0}\! d\xi\, e^{-N\xi} G(\xi,t), }  the
associated anomalous dimension
$\gamma(\as(t),N)$ corresponds  to the largest eigenvalue in the singlet
sector. 
At large $t$ and fixed $\xi$ the evolution equation  in $N$-moment space is
then
\eqn\tevol{
\frac{d}{dt}G(N,t)=\gamma(\as(t),N) G(N,t), }  where $\as(t)$ is the running
coupling. 

The perturbative anomalous dimension  is
known up to NNLO~\nnlo,  given by
\eqn\gammadef{
\gamma(\as,N)=\alpha_s \gamma_0(N)~+~\alpha_s^2
\gamma_1(N)~+~\alpha_s^3
\gamma_2(N)\dots .}
The
corresponding splitting function is
related by a Mellin  transform to $\gamma(\as,N)$:
\eqn\psasy{
\gamma(\as,N) =\int^1_0\!dx\,x^N\!P(\as,x). }  
Small $x$ for the splitting
function corresponds to small $N$ for  the
anomalous dimension: more precisely
$P\sim (1/x)(\log(1/x))^n$ corresponds to $\gamma\sim n!/N^{n+1}$.  Even assuming
that a leading twist description of scaling
violations is still valid in  some range of small $x$, as soon as
$x$ is small enough that
$\as \xi\sim 1$, with $\xi=\log{1/x}$, all terms of order
$(\as/x) (\as \xi)^n$ (LLx) and $\as(\as/x)  (\as \xi)^n$ (NLLx)  which are
present in the splitting functions must be
considered in order to achieve an accuracy up  to terms of order
$\as^2(\as/x) (\as \xi)^n$ (NNLLx).

These terms can in principle be derived from the knowledge of the kernel
$\chi(\as,M)$ of the BFKL
$\xi$--evolution equation
\eqn\xevol{
\frac{d}{d\xi}G(\xi,M)=\chi(\as,M) G(\xi,M), }  which is satisfied at large
$\xi$ by the Mellin  transform of the parton distribution
\eqn\Mmom{  G(\xi,M)=\int^{\infty}_{-\infty}\! dt\, e^{-Mt} G(\xi,t). }
The evolution  kernel in eq.~\xevol\ is the Mellin transform of the
BFKL kernel $K\left(\alpha_s,\frac{Q^2}{k^2}\right)$, which determines
the high-$s$ limit of
the cross-section for the scattering of two gluons with virtualities $k^2$ and
$Q^2$ into $n$ gluons
\eqn\bfklker{\chi(\alpha_s,M)=\int_{0}^{\infty} \frac{dQ^2}{Q^2}
\left(\frac{Q^2}{k^2}\right)^{-M} K\left(\alpha_s,\smallfrac{Q^2}{k^2}\right).
}
This kernel 
has been computed to
NLO accuracy~\refs{\fl\fleta\dd{--}\cc}
\eqn\chiexp{
\chi(\as,M)=\alpha_s \chi_0(M)~+~\alpha_s^2 \chi_1(M)~+~\dots . }
Notice that in gluon--gluon scattering $\xi=\ln(s/\sqrt{Q^2 k^2})$,
whereas in deep-inelastic scattering $\xi \sim \ln (s/Q^2)$. The kernels
that correspond to these two cases coincide at leading order, and beyond
leading order are related in a simple way, as we
shall discuss explicitly in sect.~3 below.

Quite generally, the anomalous dimension $\gamma$ eq.~\gammadef\ and
the BFKL kernel are related due to the
fact~\refs{\afp{,}\sxres{,}\sxphen}\ 
that the solutions of the BFKL and GLAP
equations coincide at leading twist if their respective evolution kernels are
related by a ``duality'' relation.  In the fixed coupling limit
the duality relations are simply given by:
\eqnn\dualdef\eqnn\dualinv
$$\eqalignno{ 
\chi(\as,\gamma(\as,N))&=N, &\dualdef\cr
\gamma(\as,\chi(\as,M))&=M. &\dualinv\cr
}
$$
This implies that
all relative corrections of order
$(\alpha_s \log{1/x})^n$ to the splitting function can be derived from
$\chi_0(M)$, those of order
$\alpha_s(\alpha_s \log{1/x})^n$ from $\chi_1(M)$ and so on: indeed,
if we expand $\gamma(\as,N)$ in powers of $\as$ at fixed
$\as/N$
\eqn\sxexp
{\gamma(\as,N)=\gamma_s(\as/N)+\as\gamma_{ss}(\as/N)+\dots,} 
eq.~\dualdef\ determines $\gamma_s$ in terms of $\chi_0$, $\gamma_{ss}$
in terms of $\chi_0$ and $\chi_1$ and so on.

Conversely, all relative corrections of order
$(\alpha_s \log{Q^2})^n$ to the BFKL kernel
$K\left(\alpha_s,\smallfrac{Q^2}{k^2}\right)$ 
eq.~\bfklker\
 can be derived from
$\gamma_0(N)$, those of order
$\alpha_s(\alpha_s \log{Q^2})^n$ from $\gamma_1(N)$ and so on: 
expanding $\chi(\as,M)$ in powers of $\as$ at fixed
$\as/M$
\eqn\sxexp
{\chi(\as,M)=\chi_s(\as/M)+\as\chi_{ss}(\as/M)+\dots,} 
eq.~\dualinv\ determines $\chi_s$ in terms of $\gamma_0$, $\chi_{ss}$
in terms of $\gamma_0$ and $\gamma_1$ and so on.

As is by now
well known, the early wisdom on how to implement the information from
$\chi_0$ was 
completely shaken by the much
softer behaviour of the data at small $x$~\refs{\das,\summ,\bfklfits} and 
by the computation of
$\chi_1$~\refs{\fl\fleta\dd{--}\cc}, which showed that the naive expansion 
for the
improved anomalous dimension had a hopelessly
bad behaviour, in particular near $M=0$ and $M=1$. In
refs.~\refs{\sxphen{,}\sxres} we have shown that the physical
requirement of momentum  
conservation  for anomalous dimensions
fixes by duality the value of $\chi(\as,M)$ at $M=0$: 
\eqn\mom{
\chi(\as,0) = 1. }
This implies that in order to cure the instability at $M=0$ 
it is sufficient to ensure that the standard
GLAP leading log resummation is performed in accordance to the
perturbative expansion of the anomalous dimension eq.~\gammadef, in
which momentum conservation is respected order by order.

The natural way to do this is by combining the
small $x$ resummation and the standard resummation of collinear
singularities. As a result one obtains a pair of `double-leading' (DL)
perturbative 
expansions of the kernel $\chi$ and anomalous dimension $\gamma$, dual
to each other order in perturbation theory (up to subleading
terms). For example, in the DL LO expression for $\chi$  an infinite
series of terms of order $(\as/M)^n$ is added to the fixed-order $\chi_0$
result, and similarly at DL NLO terms of order $\as(\as/M)^n$. The
net result is to obtain a regular function at $M=0$ which only
violates momentum by small subleading terms, and interpolates between
the BFKL result at small $x$ ($M\sim\half$) and the GLAP result at
large $x$ ($M\sim 0$). 

The DL
expansion kernels are
\eqn\kiDL{
\chi_{DL}=\chi_{DL\,LO}+\as\chi_{DL\,NLO}+\dots } 
with
\eqn\kiDLLO{
\chi_{DL\,LO}(\as,M) = \as \chi_0(M) +
\chi_s(\smallfrac{\as}{M})-\frac{n_c\as}{\pi M}}
\eqn\kiDLNLO{
\chi_{DL\,NLO}(\as,M) = \as\chi_1(M)+\chi_{ss}(\smallfrac{\as}{M})-
\as(\frac{f_2}{M^2}+\frac{f_1}{M})-f_0} 
The kernels $\chi_s$ and $\chi_{ss}$ are derived from $\gamma_0(N)$
and $\gamma_1(N)$ using the inverse duality relation 
\eqn\dualinv{\gamma(\as,\chi(\as,M))=M}
and are then given by:
\eqn\kis{\gamma_0\left(\chi_s\left(\smallfrac{\as}{M}\right)\right)=
\smallfrac{M}{\as},}
\eqn\kiss{\chi_{ss}\left(\smallfrac{\as}{M}\right)=
-\frac{\gamma_1(
\chi_s(\smallfrac{\as}{M}))}{\gamma_0'(\chi_s(\smallfrac{\as}{M}))},}
while the numbers $f_0$, $f_1$ and $f_2$ are fixed so as to remove 
double counting between $\chi_1$ and $\chi_{ss}$.

The DL kernel eq.~\kiDL\ has a stable perturbative expansion for small
$M$, and indeed for all 
$M\lsim\half$. However, it is still unstable near $M=1$:
while
momentum conservation and duality fix $\chi=1$ at $M=0$, near $M=1$
$\chi_0$ behaves like $1/(1-M)$ and $\chi_1$ behaves like
$-1/(1-M)^2$. Hence, the NLO approximation for $\chi_{DL}$
has no minimum --- in fact, the minimum is absent at any even  order
of the perturbative expansion eq.~\kiDL. This prevents a precise
treatment of non leading terms, for which a
better control of the central region near $M=1/2$ is needed,
which in turn creates a problem for the evaluation of the small
$x$ asymptotic behaviour, which depends on the central region. In
ref.~\sxphen, however, it was found that phenomenology requires
the all--order minimum of $\chi$ to significantly differ from its
low-order one, and that a  successful description of the data
is obtained if  the minimum
value of $\chi$ is taken as a free parameter to be fitted to the HERA
data. 

So far we have assumed that the running of the coupling is treated
perturbatively in the DL expansion. In particular, the running of the
coupling is a leading $\ln Q^2$, but subleading $\ln \smallfrac{1}{x}$
effect. As a consequence, when the coupling runs
the duality relation eq.~\dualdef\ is unaffected at the leading  $\ln
\smallfrac{1}{x}$ level, while at the N$^k$LL$x$ it receives a perturbatively computable
$O\left[\left(\beta_0\as\ln \smallfrac{1}{x}\right)^k\right]$
correction.
Specifically, at NLO
\eqn\rccor{\gamma_{ss}(\as/N)=-\frac{\chi_1(\gamma_s(\as/N))}
{\chi_0^\prime(\gamma_s(\as/N))}-\beta_0\frac{\chi_0^{\prime\prime}
(\gamma_s(\as/N))\chi_0(\gamma_s(\as/N))}{2[\chi_0^\prime(\gamma_s(\as/N))]^2.}}
The running coupling correction may be viewed as an effective
contribution to $\chi$: defining
\eqn\rcccor{\chi_{1\,pert.}^{\beta_0}(M)=\beta_0\frac{\chi_0^{\prime\prime}
(M)\chi_0(M)}{2\chi_0^\prime(M)}}
$\gamma_{ss}$ eq.~\rccor\ is the naive (fixed coupling) dual of an
effective running coupling $\chi$ whose NLO term is supplemented by
$\chi_{1\,pert.}^{\beta_0}$ eq.~\rccor.
However, it is apparent that this effective $\chi$ is singular as
$M\to\half$, which implies~\sxap\ that the corresponding splitting
function is  enhanced as
$x\to0$. In fact, the contribution $\Delta P_{s^n}(\as,x)$ 
to the N$^n$LLx splitting function induced
by the running-coupling  duality correction 
to the splitting function behaves as
\eqn\rccorrn{
{\Delta P_{s^n}(\as,x)\over P_s(\as,x)}\tozero{x} \left(\beta_0\as
\ln\smallfrac{1}{x}\right)^n.} 

The all-order resummation of these running
coupling small $x$ corrections is therefore needed, and it turns out~\refs{\sxrun{,}\runph}
to be, on top of momentum conservation, the second crucial ingredient
of the resummation procedure.  The running coupling  resummation can
be performed~\sxrun\ by 
taking a quadratic approximation to the BFKL kernel in the vicinity of
its minimum.
Indeed,  in $M$
space the usual running coupling $\as(t)$ becomes a differential
operator: assuming the coupling to run in the usual way with $Q^2$, and
 taking only the one-loop beta function
into account, one has
\eqn\ashdef{
\ash = \frac{\as}{1-\beta_0 \as \smallfrac{d}{dM}}, } where $\beta_0$ is
the first coefficient of the
$\beta$-function (so
$\beta=-\beta_0\as^2+\cdots$),  with
the obvious generalization to higher orders.
Eq.~\ashdef\ implies that 
\eqn\bascomm{[\ahat,M]=\beta_0\ahat^2.}
Therefore, the explicit form of a general function of $\ahat$ and $M$
will depend on the choice of operator ordering. 
The
$\xi$-evolution equation eq.~\xevol\  becomes~\Liprun
\eqn\xevolrun{
\frac{d}{d\xi}G(\xi,M)=\chi(\ash,M) G(\xi,M), }  where the derivative with
respect to $M$ acts on everything to the right,
and
$\chi$ may be expanded as in eq.~\chiexp\ keeping the powers  of $\ash$ on the
left. It is easy to show that different choices of operator ordering
in eq.~\xevolrun\ would corresponds to different choices for the
argument of the running coupling~\refs{\Liprun{,}\ciafresb}.

When $\chi$ is approximated by $\as \chi_0$,
eq.~\xevolrun\ becomes, after taking a second Mellin transform,
\eqn\asheq{
(1-\beta_0\as \frac{d}{dM})N G(N,M) + F(M) = \as \chi_0(M) G(N,M), }
where the function $F(M)$ is a boundary condition.
By expressing the anomalous dimension in terms of the exact solution
to eq.~\asheq, it is possible to prove~\sxrun\ that corrections
eq.~\rccorrn\ to
naive duality eq.~\dualdef\ which generalize the NLO term eq.~\rccor\
may be determined perturbatively, thereby ensuring that duality
continues to hold at the running coupling level to all perturbative
orders. Because the dual GLAP equation enjoys the standard
factorization properties, this also implies perturbative factorization of the
running coupling BFKL equation.
Furthermore, the all-order sum of these running coupling corrections can be
determined explicitly  
when $\chi_0(M)$ is replaced by its quadratic
approximation near its minimum at $M=1/2$: $\chi_0(M)\sim c+\half k
(M-\half)^2$, because in such case the inverse Mellin eq.~\Mmom\ of the
solution to eq.~\asheq\ can be determined exactly in terms of Airy
function  of an argument
that depends on $\as(t)$ and $N$. Since
the asymptotic small $x$ behaviour  is determined by the central
interval in $M$ of $\chi$, one can show that the quadratic expansion
leads to the correct asymptotic behaviour implied by the kernel
$\chi_0$. 
The all-order resummation of running coupling effects can then be
obtained~\sxrun\ by showing that this exact solution  based on the quadratic
approximation to $\chi$ determines the asymptotic small $x$ behaviour
of the solution computed from the full $\chi$.

The running coupling resummation of the DL anomalous
dimension is found to greatly soften the asymptotic small $x$ behaviour in
comparison to the DL result with unresummed running coupling
effects, by replacing the naive DL small $x$ intercept with an
effective one~\refs{\ciafresa,\sxrun,\runph}. As a consequence, it is 
no longer necessary to fit the
small $x$ behaviour to the data in order to obtain reasonable
phenomenology~\refs{\sxrun,\runph}.
This, however, makes a precise description of the minimum of $\chi$
mandatory. Specifically, at NLO of the DL expansion where no minimum is
present there is a loss of predictivity because it must be assumed
that the minimum is present in the all order result~\runph.
Fortunately, the minimum is restored at each order by exploiting the
underlying BFKL symmetry, as we explain in the next section. 

\newsec{Symmetrisation}

We have seen that for a precise treatment of non leading terms a
better control of the central region near $M=\half$ is needed and for
this 
one has to stabilize the bad behaviour of $\chi$ near $M=1$.
This can be done
using the symmetry of the BFKL kernel. The
underlying Feynman diagrams which determine the kernel
$K\left(\alpha_s,\smallfrac{Q^2}{k^2}\right)$ in eq.~\bfklker\ order 
by order in perturbation theory are symmetric upon 
the exchange of the incoming and outgoing gluon. But
the interchange of the incoming and outgoing gluon virtualities $Q^2
\leftrightarrow k^2$ in the argument of the BFKL kernel
$K\left(\alpha_s,\smallfrac{Q^2}{k^2}\right)$ in eq.~\bfklker\ 
corresponds to the 
transformation $M\to 1-M$ in the argument of its Mellin transform
$\chi(\as,M)$. Hence, $\chi(\as,M)$ must be symmetric upon this
interchange.

The symmetry, however, 
is broken by two effects: the running of the coupling and
the fact that in deep-inelastic scattering the $N$-Mellin transform
eq.~\Nmom\ is performed with respect to the variable $\xi=\ln (s/Q^2)$ which depends on the outgoing scale $Q^2$ only. 
We will start with the fixed $\as$ case, while
running coupling effects and the associate symmetry breaking will be
discussed in the next section. The
kernel is then only asymmetric due to the choice of definition of $\xi$.
As is well known~\fl, this asymmetry corresponds to a reshuffling of
the arguments $M$,~$N$ of $\chi$: namely, a symmetrization $\chi_\Sigma$ of 
the kernel $\chi$ of eq.~\xevolrun\ can be obtained from
a symmetric kernel $\chi_{\sigma}$ which corresponds to the
symmetric choice $\xi=\ln(s/\sqrt{Q^2 k^2})$ through the implicit equation~\fl:
\eqn\symm{
\chi_\Sigma(\as,M+\half\chi_{\rm \sigma}(M))=\chi_{\sigma}(\as,M).}

Hence, we can implement the symmetrization, thereby resumming 
the $M=1$ poles, by performing the double-leading
resummation of $M=0$ poles of $\chi$, determining the
associated $\chi_\sigma$ through eq.~\symm, then symmetrizing it (as
$\chi_\sigma$ must be symmetric), and finally going back to DIS
variables by using eq.~\symm\ again in reverse.

The procedure can be understood most easily by first
considering a toy case, where we pretend that the anomalous
dimension exactly coincides with its leading--order form $\as \gamma_0(N)$, so that the BFKL kernel is just
given by $\chi_s(\as/M)$ from eq.~\kis. This corresponds to a leading $\ln
Q^2$ (or collinear)  approximation to the BFKL kernel. Symmetrization is enforced by using
eq.~\symm\ to relate the DIS kernel to the symmetric one:
we replace $M$ by $M+N/2$ and
symmetrize for $M\to (1-M)$. Note that this is allowed at the leading
$\ln Q^2$ level, because the symmetrizing terms are by construction
free of poles at $M=0$ and are thus subleading in $\ln Q^2$.

 We thus obtain the following pair of kernels, related by eq.~\symm:
\eqnn\sigtoysimm\eqnn\sigtoyDIS
$$\eqalignno{\bar\chi_{\sigma\,toy}\left(\as,M,N\right) &=
\chi_s\Big(\smallfrac{\as}{M+\smallfrac{N}{2}}\Big)+ 
\chi_s\Big(\smallfrac{\as}{1-M+\smallfrac{N}{2}}\Big)&\sigtoysimm\cr
\bar \chi_{\Sigma\,toy}\left(\as,M,N\right) &= \chi_s\left(\smallfrac{\as}{M}\right)+
\chi_s\left(\smallfrac{\as}{1-M+N}\right),&\sigtoyDIS\cr}$$ 
where $\bar\chi_{\sigma\,toy}$ is symmetric, 
and $\bar\chi_{\Sigma\,toy}(\as,M)$ is
relevant for DIS, i.e. dual to the DIS anomalous dimension. 

Equations~\sigtoysimm, \sigtoyDIS\ should  be viewed as implicit
definitions of the corresponding kernels $\chi_\sigma(\as,M)$, 
$\chi_\Sigma(\as,M)$ 
as solutions of equations of the form
\eqn\expsimm{\chi(\as,M)=\bar\chi\left(\as,M,\chi(\as,M)
\right)}
in respectively symmetric or asymmetric variables. So $\bar\chi(\as,M,N)$ 
can be viewed as``off-shell'' continuations of the kernels
$\chi(\as,M)$, which are obtained from them by putting 
$N$ ``on-shell'' (see eq.~\dualdef).  Alternatively we can use 
use the off-shell kernels as implicit definitions of the anomalous 
dimensions $\gamma(\as,N)$ through the duality condition \dualinv :
\eqn\expsimn{N =\bar\chi\left(\as,\gamma(\as,N),N
\right).}
Summarising, we obtain $\chi(\as,M)$ or $\gamma(\as,N)$ by putting 
either $N$ or $M$ ``on-shell'' (i.e. set $N=\chi(\as,M)$ or 
$M=\gamma(\as,N)$ respectively) in the 
``off-shell'' relation $N=\bar\chi(\as,M,N)$.

Note that as $N\to\infty$, the symmetrising contribution in eq.~\sigtoyDIS\ 
$\chi_s(\smallfrac{\as}{1-M+N})\sim \chi_s(\smallfrac{\as}{N})
\sim O(\smallfrac{\as}{N})$. Thus this term does not spoil the 
identification of the resummed kernel $\chi$ with the GLAP result
$\chi_s$ at large $N$, which in turn guarantees that 
the resulting anomalous dimension
matches smoothly to the GLAP anomalous dimension.

However the symmetrizing contribution does violate the momentum
conservation condition $\chi_{\Sigma\,toy}(\as,0)=1$, albeit by subleading
terms. Indeed,  at the momentum conservation point $M=0$, $\chi_s(\as/M)$
by construction satisfies $\chi_s(\as/M)|_{M=0}=1$. However,
$\chi_s(\as/(1-M))|_{M=0}=\chi_s(\as)=O(\as)$, and
$\chi_s(\as/(1-M+N))|_{M=0}= \chi_s(\as)(1+ O(N))=
O(\as)$, where the last step follows from duality
eq.~\dualdef.  Hence, the momentum violation is $O(\as)=O(\as^{n+1} M^{-n})$,
or next-to-leading order in an
expansion  of $\chi$ in powers of $\as$ at fixed $\as/M$.
A simple way of enforcing momentum conservation without spoiling the symmetrization 
properties is then to add to eqs.~\sigtoysimm,\sigtoyDIS\ an extra term of the form
\eqn\momdef
{\chi_{\rm mom}(\as,N)=c_m f_m(N),
}
where $f_m(1)=1$ and $f_m(0)=f_m(\infty)=0$, for example
\eqn\fdef
{f_m(N)=\frac{4N}{(N+1)^2},} 
and with the constant $c_m$  is chosen to fix up the momentum constraint.
The condition $f_m(0)=0$ ensures that $\chi_{\rm mom}(\as,N)$ is subleading:
because $c_m=O(\as)$,
$\chi_{\rm mom}(\as,N)$ eq.~\momdef\ is of order $O(\as N) \left[1+ O(\as)+
O(N)\right]$. But by the duality eq.~\dualdef\ $N=O(\as/M)$, hence 
$\chi_{\rm mom}(\as,N) $ is $O(\as^2 /M)$, i.e. next-to-leading 
$\ln Q^2$. The condition $\lim_{N\to\infty}f_m(N)=0$ 
again ensures that the extra
term does not spoil the indentification of the resummed
$\chi$ kernel with the GLAP result $\chi_s$ at large $N$.

The full symmetrized double--leading result can be obtained through a
similar procedure, but starting with
$\chi_{DL\,LO}(M)$, given in eq.~\kiDLLO. Clearly, the contribution from
the leading--order BFKL kernel $\chi_0$ is already symmetric, but the
symmetry should only hold in symmetric variables, while in DIS
variables it must be asymmetric. Furthermore the collinear and anti-collinear 
singularities must match exactly with those derived from GLAP.
To achieve this it is necessary to modify the BFKL kernel by 
subleading terms in the same way that we symmetrised the pure 
GLAP model above. Recall that the LO BFKL kernel
\eqn\kizero{
\chi_0(M) = \smallfrac{n_c}{\pi}\left[2\psi(1)-\psi(M)-\psi(1-M)\right],}
where 
$\psi(M)\equiv \smallfrac{d}{dx} \ln \Gamma(x)$ is the polygamma
function. We can separate it into a sum of two pieces: a collinear piece 
$\chi_0^+$ with poles at $M=0,-1,-2,\ldots$ 
coming from the collinear region $Q^2>k^2$
and an anticollinear piece $\chi_0^-$ 
with poles at $M=1,2,3,\ldots$ coming from the 
anti-collinear region $Q^2<k^2$: $\chi_0=\chi_0^+ + \chi_0^-$ with
\eqn\chicoll{\eqalign{
\chi_0^+(M)&=\frac{n_c}{\pi}\left(\psi(1)-\psi(M)\right)\cr
\chi_0^-(M)&=\chi_0^+(1-M).\cr}}
It follows that in symmetric variables the off-shell 
extension of $\chi_0$ must be
\eqn\kizerosym{
\bar\chi_0(M,N) = \smallfrac{n_c}{\pi}\left[\psi(1)+\psi(1+N)-\psi(M+N/2)-
\psi(1-M+N/2)\right],}
so that in DIS variables 
\eqn\kizeroDIS{
\bar\chi_0(M-\smallfrac{N}{2},N) 
= \smallfrac{n_c}{\pi}\left[\psi(1)+\psi(1+N)-\psi(M)-
\psi(1-M+N)\right].}
The leading collinear and anticollinear poles
\eqn\kizerodc{
\bar\chi^{dc}_0(M,N) = \smallfrac{n_c}{\pi}\left[\smallfrac{1}{M+N/2}
+\smallfrac{1}{1-M+N/2}\right],}
then coincide with those from GLAP  [eq.~\sigtoysimm\ expanded to $O(\as)$], 
while at the same time the only difference between \kizerosym\ and \kizero\
is in terms of $O(N)=O(\as)$ [using \dualdef\ with $\chi(M,\as)
=\as\chi_0(M)$]. Note that besides the shift in the argument of  
$\psi(1-M)$ we also added a term to $\chi_0^-$ proportional to 
$\psi(1+N)-\psi(1)$: this is to ensure that at large $N$ in DIS variables the 
anticollinear terms $\psi(1+N)-\psi(1-M+N)\sim O(1/N)$, and thus that the
matching to GLAP in the collinear region (and in particular at large $N$)
is not spoiled. 

Putting together the GLAP and BFKL components, the symmetrized 
off-shell leading order DL $\chi$ is then
\eqn\simmLO{\bar\chi_{\sigma LO}(\as,M,N) =
\chi_s\left(\smallfrac{\as}{M+\smallfrac{N}{2}}\right)
+ \chi_s\left(\smallfrac{\as}{1-M+\smallfrac{N}{2}}\right)
+\as\tilde\chi_0(M,N)+\chi_{\rm mom}(\as,N),}
where if
\eqn\chicollos{\eqalign{
\bar\chi_0(M,N)&=\bar\chi^+_0(M,N)+\bar\chi^-_0(M,N),\cr
\bar\chi^+_0(M,N)&= \smallfrac{n_c}{\pi}(\psi(1)
-\psi(M+\smallfrac{N}{2})),\cr
\bar\chi^-_0(M,N)&= \smallfrac{n_c}{\pi}(\psi(1+N)
-\psi(1-M+\smallfrac{N}{2})),\cr}}
then
\eqn\chicollosx{\eqalign{
\tilde\chi_0(M,N)&=\bar\chi_0(M,N)-\bar\chi^{dc}_0(M,N)\cr
&=\smallfrac{n_c}{\pi}\left(\psi(1)+\psi(1+N)
-\psi(1+M+\smallfrac{N}{2})-\psi(2-M+\smallfrac{N}{2})\right),
}}
where in the last step we have used the identity $\psi(x+1)=\psi(x)
+\smallfrac{1}{x}$. In DIS variables we get
\eqn\sigLODIS{
\bar\chi_{\Sigma\,LO}(\as,M,N) = 
\chi_s\left(\smallfrac{\as}{M}\right)+
\chi_s\left(\smallfrac{\as}{1-M+N}\right)+
\as\tilde\chi_0(M-\smallfrac{N}{2},N)+\chi_{\rm mom}(\as,N),}
where 
\eqn\chicollosdis{
\tilde\chi_0(M-\smallfrac{N}{2},N)=
\smallfrac{n_c}{\pi}\left(\psi(1)+\psi(1+N)
-\psi(1+M)-\psi(2-M+N)\right).
}
This differs from the original unsymmetrized leading order DL result
eq.~\kiDLLO\ by subleading terms, as we now show. The symmetrizing contribution
$\chi_s[\as/(1-M)]$ is $O(\as)$ and free of $M=0$ poles, hence
next-to-leading $\ln Q^2$. Because it is an $O(\as)$ contribution to
$\chi$ it is leading $\lnx$. However, its $O(\as)$
term is subtracted by the double counting term 
$-\smallfrac{\as n_c}{\pi}\smallfrac{1}{1-M+N}$, and
what remains is $O(\as^2)$, namely next-to-leading
$\lnx$. Finally, the leading $\lnx$ contribution to
$\chi_{\Sigma\,LO}$ eq.~\sigLODIS\ differs from $\chi_0$ because of
the $N$-dependent shift of the argument $M$: however, by duality
eq.~\dualdef\  $N=O(\as)$ and thus the difference is $O(\as^2)$, namely
next-to-leading $\lnx$.
The violation of momentum conservation in the absence of $\chi_{\rm mom}$
is still subleading. This is proven by simply
observing  that 
$\chi_{\rm mom}(\as,N)$  differs from the DL result by subleading
terms. But the DL result respects momentum conservation up to
subleading terms, and the remaining terms are subleading with respect
to it everywhere. 

In a similar way, starting from $\chi_{DL\,NLO}(M,\as)$ in eq.~\kiDLNLO\ ,
one can write down the symmetrized off-shell NLO DL $\chi$: 
\eqn\simmNLO{
\bar\chi_{\sigma NLO}(\as,M,N) =
\chi_{ss}\left(\smallfrac{\as}{M+\smallfrac{N}{2}}\right)
+ \chi_{ss}\left(\smallfrac{\as}{1-M+\smallfrac{N}{2}}\right)
+\as\tilde\chi_1\left(M,N\right)+\chi_{\rm  mom}(\as,N).}
In DIS variables this corresponds to:
\eqn\sigNLODIS{
\bar\chi_{\Sigma\,NLO}(\as,M,N) =
\chi_{ss}\left(\smallfrac{\as}{M}\right)+
\chi_{ss}\left(\smallfrac{\as}{1-M+N}\right)
+\as\tilde\chi_1\left(M-\smallfrac{N}{2},N\right)
+\chi_{\rm mom}(\as,N).}
The function $\tilde\chi_1\left(M,N\right)$, which
is related to the Fadin--Lipatov kernel $\chi_1$ (the next-to-leading
contribution to the BFKL kernel eq.~\chiexp) minus double counting, up to 
subleading corrections due to the symmetrization and to the running
of the coupling, will be constructed explicitly in section~6 below.

Let us now turn to the generic properties of the symmetrized on-shell kernels,
obtained by solving eq.~\expsimm\ in either symmetric or DIS variables.
First, it is clear that, because  momentum conservation in asymmetric
(DIS) variables fixes $\chi(\as,0)=1$, in symmetric variables the
associate symmetric kernel will satisfy the constraint
\eqn\momsym{
\chi_\sigma\left(\as,-\smallfrac{1}{2}\right)=\chi_\sigma\left(\as,\smallfrac{3}{2}\right)=
1.}    
When transforming back to DIS variables, this will lead to
\eqn\momDIS{
\chi_\Sigma(\as,0)=\chi_\Sigma(\as,2)=1.}
Since in the central region $M\sim\half$, $\chi\sim O(\as)$ and thus rather 
less than one, at least for small enough values of $\as$, on the real axis 
of the $M$-plane the symmetrized kernel will
have a minimum order by order in the (symmetrised) double leading expansion.
Since the change of variables from symmetric to DIS variables is just a shift 
$M\to M-\smallfrac{N}{2}$, there will be a minimum is DIS variables too. This
is an important generic result, since it implies necessarily that at fixed 
coupling the splitting function must behave as a power of $x$, order by order 
in resummed perturbation theory.

Now consider the position and curvature around the minimum. While
$\chi_{\sigma}$ has the minimum at $M=\half$, the minimum of
$\chi_\Sigma$ is displaced, but the value at the minimum and the 
curvature are the same. To see this, we rewrite the implicit
equation eq.~\symm\ as 
\eqn\symnonsym{
\chi_{\sigma}(\as,M- \half \chi_\Sigma(M))= \chi_\Sigma(\as,M),}
where $\chi_{\sigma}(\as,M)=\chi_{\sigma}(\as,1-M)$ is stationary at $M=\half$.
Eq.~\symnonsym\ implies that the derivatives of $\chi_\Sigma$ and
$\chi_{\sigma}$ are related by 
\eqn\dsymnonsym{\eqalign{
\chi^\prime_\Sigma(\as,M)
&= {{\chi_{\sigma}}^\prime(\as,M-\half \chi_\Sigma(M)) \over 
1+\half\chi_\sigma^\prime(\as,M-\half \chi_\Sigma(M))},\cr
\chi^{\prime\prime}_\Sigma(\as,M)
&= {{\chi_{\sigma}}^{\prime\prime}(\as,M-\half \chi_\Sigma(M)) \over 
(1+\half\chi_\sigma^\prime(\as,M-\half \chi_\Sigma(M)))^3}, }}
where the prime denotes differentiation with respect to $M$.
It follows that the minimum of $\chi_\Sigma(\as,M)$ is at $M_s$, determined by
\eqn\mrel{M_s= \half+\half \chi_\Sigma(\as,M_s),}
which implies
\eqn\ckrel{\chi_{\sigma}(\as,\half)=\chi_\Sigma(\as,M_s),\qquad
\chi''_{\sigma}(\as,\half)
=\chi''_\Sigma(\as,M_s),}
i.e., the intercept and curvature of $\chi_\Sigma$ and $\chi_\sigma$ at the
respective minima are the same (note that this is not true for higher
derivatives of either function:  after all, $\chi_\Sigma$ is asymmetric). 
Hence, whereas the location $M_s$ of the minimum is shifted away from
$M_s=\half$  by the asymmetric scale choice, the intercept and
curvature at the minimum are unaffected.

\topinsert
\vbox{
\epsfxsize=12truecm
\centerline{\epsfbox{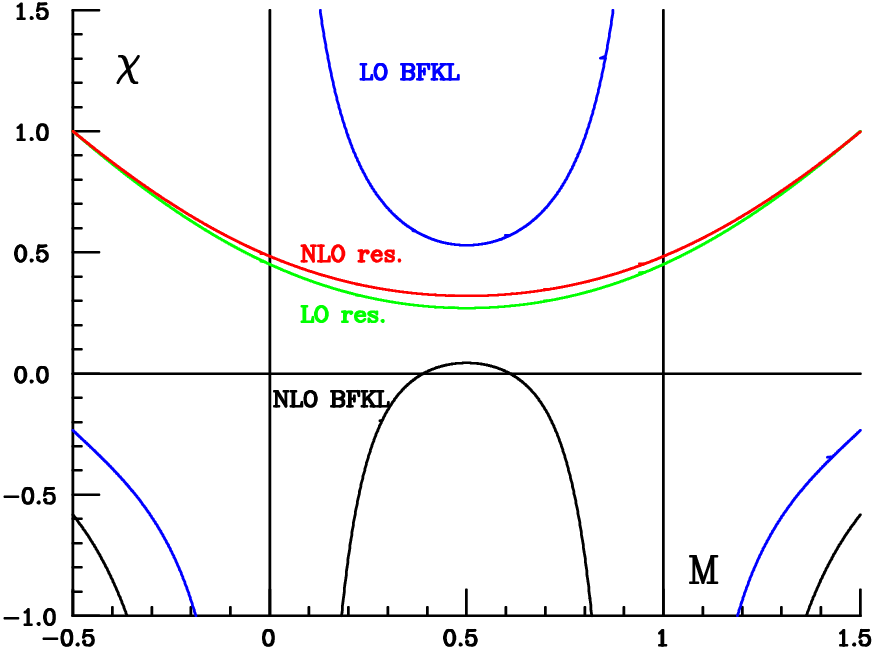}}
\bigskip
\hbox{
\vbox{\footnotefont\baselineskip6pt\narrower\noindent Figure 1: Plot
of different symmetric small $x$ kernels $\chi$. From the top (in the
central region): the LO BFKL kernel $\as \chi_0$, the NLO and LO
resummed DL expansion kernels $\chi_\sigma$, both on-shell and
expressed in symmetric variables, the NLO BFKL kernel $\as \chi_0+\as^2
\chi_1$. All curves are determined with $\beta_0=0$ (fixed coupling),
$\as=0.2$ and $n_f=0$. Note
the unphysical branches of  the LO
and NLO BFKL curves, outside the interval
$0<M<1$.  
 }}\hskip1truecm}
\endinsert 

\topinsert
\vbox{
\epsfxsize=12truecm
\centerline{\epsfbox{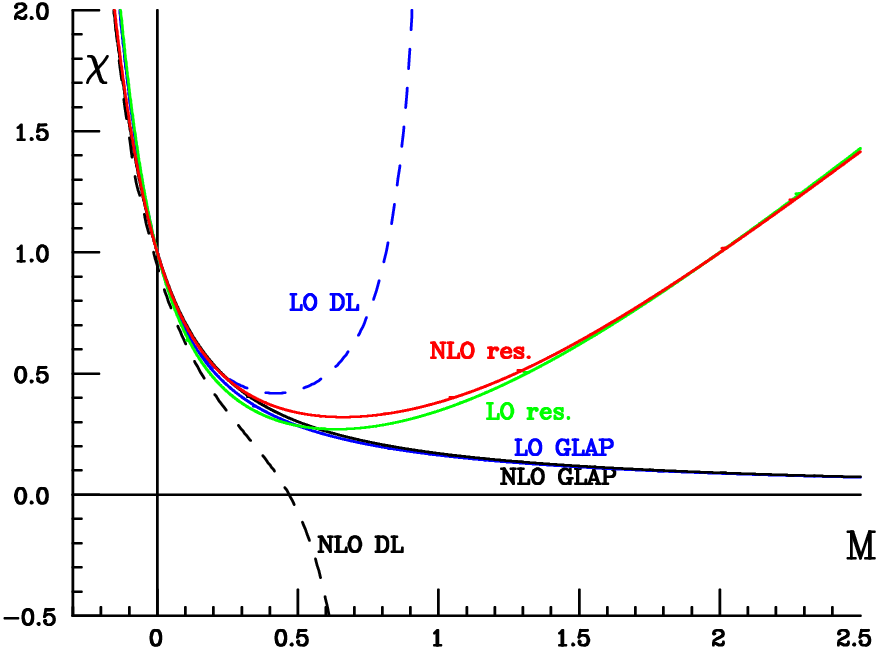}}
\bigskip
\hbox{
\vbox{\footnotefont\baselineskip6pt\narrower\noindent Figure 2: 
The resummed kernels of fig.~1 expressed in
asymmetric variables as relevant for the DIS structure function
evolution. Also shown for comparison the LO and NLO GLAP kernels and
the LO and NLO double-leading (DL) kernels. 
 }}\hskip1truecm}
\endinsert 

Although $\chi_{DL}$ has poles at $M = \pm 1,\pm 2,\ldots$, in symmetric 
variables $\chi_\sigma$ is actually an entire function of M. To see this, 
note that the off-shell $\bar\chi_\sigma(\as,M,N)$ is of the form 
$\varphi(M+\smallfrac{N}{2}) + \varphi(1-M+\smallfrac{N}{2})$, where 
$\varphi(M)$ has poles on the negative real axis at $M = -n$, 
$n = 1,2,\ldots$,  so $\bar\chi_\sigma(\as,M,N)$ will have poles at
$M+\smallfrac{N}{2}=-n$, near which it must go to infinity. Eq.~\expsimm\ 
then shows that $N\to\infty$ there, i.e.,  as we go on-shell the corresponding
singularity in $\chi_\sigma(\as,M)$ is shifted to the point at infinity. 
Similarly, in DIS variables $\bar\chi_\Sigma(\as,M,N)$ is of the form 
$\varphi(M) + \varphi(1-M+N)$, so by a similar argument the poles on the 
positive real axis are shifted to infinity when we go on-shell, and 
$\chi_\Sigma(\as,M)$ is free of singularities to the right of $Re M = -1$.
The singularities at $M = -1, -2,\ldots$ correspond of course 
to higher twists, which are outside the control of our approximations,
but die off rapidly at high $Q^2$.

Note that whereas $\chi_\sigma(M)$ is an entire function, its 
perturbative expansion still has order by order singularities at 
$M=0$, just like $\chi_{DL}(\as,M)$ and $\chi_\Sigma(\as,M)$. 
In fact whereas the perturbative expansions $\chi_{DL}(\as,M)$ and 
$\chi_\Sigma(\as,M)$ have single collinear poles (of the form $(\as/M)^n)$ 
the perturbative expansion of $\chi_\sigma(\as,M)$
has double singularities of the form $\as^n/M^{2n-1}$. 
To see this substitute the expansion eq.~\chiexp\ in eq.~\symm\ to get
\eqn\ptsymnonsym{{\chi_\Sigma}_0(M)={\chi_\sigma}_0(M),
\quad{\chi_\Sigma}_1(M)={\chi_\sigma}_1(M)-\half \chi_0(M)\chi_0'(M), 
\ldots,}
whence the result. Similarly the anticollinear singularities in 
DIS variables, i.e in $\chi_\Sigma(\as,M)$ at $M=1$ are 
also double singularities, of the form $\as^n/(1-M)^{2n-1}$.  
The
spurious cubic singularity in $\chi_1$ is well
known~\refs{\fl{,}\ciaf} and will be discussed in section~5 when
extending the symmetrization to next-to-leading order.

A comparison of the unresummed  BFKL kernels $\as\chi_0$ and
$\as\chi_0+\as^2\chi_1$ with the symmetrised DL
expansions at LO and NLO is presented in fig.~1. The curves shown
are, in symmetric variables, the on-shell versions obtained by using
eq.~\expsimm, from eqs.~\simmLO,~\simmNLO. First, we observe that the
LO and NLO symmetrised DL curves are very close, indicating a good behaviour
of the corresponding expansion, while the NLO BFKL kernel was in all
respects completely different from its LO approximation. 
We also observe that the DL expressions are very smooth and 
rather flat. This is due to the fact that the symmetrization procedure 
shifts the poles at infinity while momentum conservation imposes
$\chi(-\half)=\chi(\smallfrac{3}{2})= 1$. In the following 
section we will use a quadratic approximation of the kernels near
their minimum. Clearly for the DL kernels the validity of this
approximation is quite accurate in the whole physical region
$0<M<1$. The resummation considerably
reduces the value of $\chi$ at the minimum in comparison to the 
LO BFKL kernel, implying a softer small $x$ asymptotic behaviour.  
Amusingly, after resummation the NLO correction is actually small 
and positive, in complete contrast to the large negative correction 
suggested by a naive interpretation of NLO BFKL. 

In fig.~2 we display a set of $\chi$ kernels expressed in
asymmetric variables as relevant for DIS structure function
evolution. The LO and NLO BFKL kernels, the LO and NLO DL curves and
finally those derived at LO and NLO from the symmetrization procedure
are compared with the dual of LO and NLO GLAP. The stabilization of
the resummed expansion for $M\gsim \half$ through the symmetrisation 
procedure is apparent: in fact it is guaranteed by the anticollinear momentum 
conservation constraint at $M=2$. The position of the minimum is shifted 
slightly to the right or $M=\half$ but its height and curvature are the 
same as in the resummed curves of fig.~1. However the resummed result 
is now very close to the GLAP result for all $M$ below the minimum of the 
resummed symmetrised curve. As we shall see in the next section, running 
coupling effects extend this agreement considerably. 

Let us finally discuss the effect of the switch from symmetric to
asymmetric variables on the expansion of the anomalous dimension.
At the fixed-coupling level, this is simply
determined using
duality eq~\dualdef\ in  eq.~\symnonsym: if $\gamma_\sigma$ and 
$\gamma_\Sigma$ are
respectively dual to the symmetric and nonsymmetric
$\chi_\sigma$ and $\chi_\Sigma$, then, to all perturbative orders 
\eqn\gsymnonsym{\gamma_\Sigma(\as,N)=\gamma_\sigma(\as,N)+\half N.}
This shows that the effect of the scale choice is subleading, 
i.e. $O(\as^{k+1} N^{-k})$. If $\gamma_\sigma$ is expanded 
according to eq.~\sxexp, then the term
induced by the scale choice (the last
term in eq.~\gsymnonsym) is a contribution to $\gamma_{ss}$ of order 
$(\as/N)^{-1}$.  Of course, if $\gamma$  is perturbative (i.e. if it
vanishes in the $\as\to0$ limit), this term will cancel
against an equal and opposite $\half N$
contribution to ${\gamma_\sigma}_{ss}$. Furthermore, an
important implication of eq.~\gsymnonsym\ is that the singularities of 
$\gamma(\as,N)$ in the complex $N$-plane are the same in both 
symmetric and asymmetric variables, and thus in particular the 
small-$N$ singularity which determines the small $x$ behaviour of the 
splitting function will also be the same.

\newsec{Running coupling}

So far, we have discussed the improvement of the $\chi$ kernel and its
dual anomalous dimension at the fixed coupling level. As
discussed in the introduction, the
main motivation for the symmetrization of the previous
section is that the symmetrized double-leading result has a stable minimum 
order by order in the resummed expansion, in contrast to  the BFKl 
or unsymmetrized DL expansions which have an instability. The
existence of a minimum enables a running coupling resummation, which
is required~\refs{\sxrun{,}\runph} if we wish to get
a stable 
resummed anomalous dimension at small $N$, and thus a uniformly stable 
splitting function at small $x$. To this purpose, in line with the 
strategy devised in refs.~\refs{\sxrun{,}\runph}, the running coupling 
evolution equation~\xevolrun\ is  solved with a quadratic approximation 
to the kernel near the minimum, but with the running of the coupling treated
exactly at the leading $\ln Q^2$ level. 
The solution is then used to improve the anomalous dimension
computed within a perturbative treatment of the running of the
coupling. 

In refs.~\refs{\sxrun{,}\runph} the quadratic approximation near the
minimum was applied to the LO BFKL kernel $\chi_0$, eq.~\kizero. This
kernel is symmetric and has a minimum at $M=\half$. However, we have
seen that in order to preserve the symmetry and the existence of the
minimum beyond the LO, we have to construct a different expansion for
$\chi$ which at LO is given by $\chi_{\sigma\,LO}(\as,M)$ obtained
from its off-shell version eq.~\sigLODIS~ through eq.~\expsimm. 

We are thus led to solve the equation
\eqn\rcee{
N G(N,M) = \chi^q_{\sigma}(\ahat,M) G(N,M)+ F(M), }
where $\chi_{\sigma}(\ahat,M)$ has been replaced by its quadratic
approximation $\chi^q_{\sigma}(\ahat,M)$ near the minimum 
\eqn\finquad{\chi^q_{\sigma}(\as,M)=c(\as) 
+\half \kappa(\as) \left(M-\half\right)^2}
and $F(M)$ is a boundary condition to
perturbative evolution. Note that, in practice,  because of
eq.~\ckrel, the intercept $c(\as)$ and curvature $\kappa(\as)$ can be computed
using the symmetric variable resummation $\chi_\sigma$. The quality of
the quadratic approximation, especially in the relevant region
$0< M <\half$, can be appreciated from fig.~3, where the resummed curves
obtained from the symmetrisation procedure, shown in fig.~1, are
compared to their quadratic approximations. 

\topinsert
\vbox{
\epsfxsize=12truecm
\centerline{\epsfbox{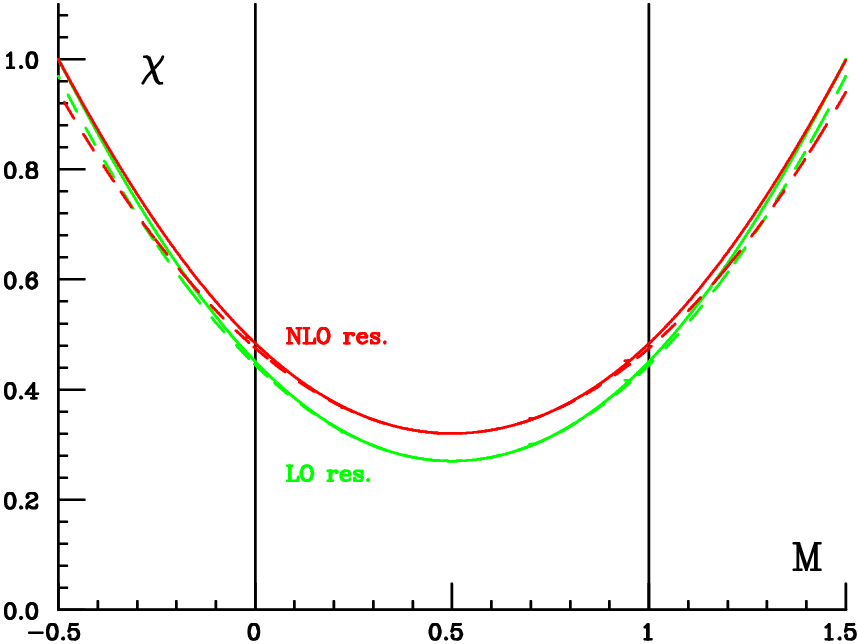}}
\bigskip
\hbox{
\vbox{\footnotefont\baselineskip6pt\narrower\noindent Figure 3: Plot
of  the NLO and LO resummed DL expansion kernels $\chi_\sigma$, both
on-shell and expressed in symmetric variables, compared with their
quadratic approximations (dashed curves) near the minimum at $M=\half$
}}\hskip1truecm} 
\endinsert 

There are two complicating features in the present case with respect
to our previous work~\sxrun. First, in earlier papers the curvature $c$ and 
intercept $\kappa$ of the kernel were linear in $\as$;  
eq.~\rcee\ could then be solved to yield a solution
expressed in terms of Airy functions. Here we need to 
expand $c(\ahat)$ and $\kappa(\ahat)$ in 
powers of $\ahat-\as$ and then, to the required accuracy, it turns 
out that the solution can be expressed in terms of Bateman functions. 
The second complication is that, since $\ahat$ is an operator which 
does not commute with $M$, any expression like $\chi(\ahat, M)$  is 
meaningful only after specifying the operator ordering. This amounts to 
specifying the argument of the running coupling. 
The first issue will be dealt with 
immediately, while the ordering issue will be postponed to the 
next two sections, where we will
explicitly construct the resummed kernels, anomalous dimensions and
splitting functions in the LO
and NLO approximations. 

When, as in the present case, $c(\ahat)$ and
$\kappa(\ahat)$ are given as power series in $\ahat$, the differential 
equation~\rcee\ cannot be solved in general. However, we can determine the
solution by expanding $\ahat$ about $\as$ to first order in
$\ahat\beta_0$. It is easy to see that this procedure guarantees
that the whole sequence of leading log contributions to $\as$ is
correctly included, up to subleading corrections. We have 
\eqnn\cexp\eqnn\kaexp
$$\eqalignno{ c(\ahat)&= c(\as)+\left(\ahat-\as\right)
c^\prime(\as)+O\left[\left(\beta_0\smallfrac{d}{dM}\right)^2
\right]&\cexp\cr
\kappa(\ahat)&= \kappa(\as)+\left(\ahat-\as\right)
\kappa^\prime(\as)+O\left[\left(\beta_0\smallfrac{d}{dM}\right)^2 \right],&\kaexp
}
$$
and, from eq.~\rcee,
\eqn\rceeexp{
\left[N -\bar c(\as) -\half\bar\kappa(\as) \left(M-\half\right)^2 \right] 
G(N,M) = \ahat \left(c^\prime(\as)+\half\kappa^\prime(\as)
\left(M-\half\right)^2 \right)G(N,M)+ F(M), }
where we have defined
\eqnn\cbardef\eqnn\kbardef
$$\eqalignno{ \bar c(\as)&\equiv c(\as)-\as c^\prime(\as)&\cbardef \cr
\bar \kappa(\as)&\equiv \kappa(\as)-\as \kappa^\prime(\as).&\kbardef\cr
}
$$
Equation~\rceeexp\ can be simplified defining $\tilde
G(N,M)$ through the implicit equation
\eqn\gprime{ G(M,N)\equiv \frac{N}{N-\bar c(\as)-\half \bar
\kappa(\as) \left(M-\half\right)^2} \tilde G(M,N),}
so $\tilde G(N,M)$ satisfies
\eqn\rceqgtil{ N  \tilde G(N,M)=\ahat \phi(M,N) \tilde  G(N,M)+ F(M),}
with the kernel
\eqn\phidef{\varphi(N,M)=\frac{N(c^\prime(\as)+\half\kappa^\prime(\as)
\left(M-\half\right)^2)}{N-\left(\bar c(\as)+\half\bar \kappa\left(M-\half\right)^2 \right)}.}

Equation~\phidef\ has the form of a leading-order $M,N$ space running
coupling BFKL
equation, generalized to the case in which the kernel $\phi(M,N)$
depends on both $M$ and $N$, and not just on $M$ as the BFKL kernel
$\chi_0$ does. As discussed in ref.~\runph, the anomalous dimension
is entirely determined by the $Q^2$ dependence of the inverse
$M$--Mellin transform of the inhomogeneous solution to this
equation, viewed as a differential equation in $M$. Furthermore, this
inhomogeneous solution to all perturbative orders satisfies a
factorization property, whereby it can be written as the solution to
the associate homogenous equation, times an $M$--independent, but
$N$--dependent boundary condition. The anomalous dimension is
therefore entirely determined by the solution to the associated
homogeneous equation. The fact that the kernel now depends also on $N$
does not affect the form of the solution, since the inhomogeneous
equation is a differential equation in $M$. Hence, using the general
solution of ref.~\sxrun\ (given as eq.~(3.21) of that
reference) we get
\eqn\gensol{\eqalign{\tilde G(N,t)&=\int_{\half-i\infty}^{\half+i\infty}
\frac{d M}{2\pi i}\exp\left[\frac{M}{\beta_0\as(t)}-\frac{1}{\beta_0}
\int_{\half}^M dM^\prime \phi(M^\prime,N)
\right]\cr
&= e^{1/2\beta_0\as(t)} \int_{-i\infty}^{+i\infty}\frac{d m}{2\pi
i} \left(\frac{1 + A(\as,N) m}{1 - A(\as,N) m}\right)^{B(\as,N)}
e^{m/\beta_0\bar\as(t)},
}}
where we have defined
\eqnn\adef\eqnn\bdef
$$\eqalignno{ A(\as,N)=&\sqrt{\frac{\half\bar\kappa(\as)}
{\left(N-\bar c(\as)\right)}}, &\adef\cr
B(\as,N)=&\frac{1}{2\beta_0}\left( \frac{c^\prime(\as)}{N-\bar
c(\as)}+\frac{\kappa^\prime(\as)}{\bar\kappa(\as)} 
\right)\sqrt{\frac{N-\bar c(\as)}{\half\bar\kappa(\as)}}, &\bdef\cr}$$
while
\eqn\abardef{\frac{1}{\bar\as(t)}
=\frac{1}{\as(t)} -\frac{\kappa^\prime(\as)}{\kappa(\as)}.}

The Mellin inversion integral can now be computed exactly, with the
result
\eqn\bate{ \tilde G(N,t)=\beta_0\bar\as(t) B(\as,N)
K_{2B(\as,N)}\left(1/\beta_0\bar\as(t) A(\as,N)\right)}
where $K_\nu(x)$ is the Bateman function, defined as the solution of the
differential equation
\eqn\bateq{- K^{\prime\prime}_\nu(x)+\left(1-\frac{\nu}{x}\right)K_\nu=0,
}
with boundary condition $K_\nu(0)=1$. Note that the argument of 
$A$ and $B$ eqs.~\adef-\bdef\ is the fixed
coupling $\as$, not $\as(t)$. This is due to the fact that in the
approximation eq.~\cexp,\kaexp, all functions of $\ahat$ are linearized
by expanding about the fixed coupling $\as$. 

The anomalous dimension is determined by taking  the logarithmic derivative of
$\tilde G(N,t)$, and then setting $\as(t)=\as$. The extra factor which relates
$\tilde G$ to $G$ eq.~\gprime\ only affects the boundary condition
which, as proven in ref.~\sxrun, does not contribute to the anomalous
dimension. Therefore, we get  
\eqn\batandim{\eqalign{\gamma_B(\as,N)&=\frac{\partial}{\partial t}
\ln \tilde G(N,t)\Big\vert_{\as(t)=\as}\cr
&= \half-\beta_0\bar\as+\frac{1}{A(\as,N)}
\frac{K_{2B(\as,N)}^\prime(1/\beta_0\bar\as A(\as,N))}
{K_{2B(\as,N)}(1/\beta_0\bar\as A(\as,N))}.
}}

The Bateman anomalous dimension can be matched to the naive
(i.e. fixed coupling) dual of 
the $\chi$ with the same procedure which was used in ref.~\sxrun\ for the
Airy anomalous dimension. Namely, we determine
$\gamma_B(\as,N)$ in the $\beta_0\to 0$ limit by
saddle-point evaluation of the 
Mellin integral eq.~\gensol.  
The leading order saddle condition is
\eqn\losad{N=\as(t)\varphi(N,M_s),}
where $\varphi(N,M)$ is given by eq.~\phidef, which reduces to
\eqn\relosad{N=c_s(\as(t),\as)+\half \kappa_s(\as(t),\as) (M_s-\half)^2,}
where 
\eqnn\csdef\eqnn\ksdef
$$\eqalignno{
c_s(\as(t),\as)&=\as(t)c^\prime(\as) +\bar c(\as)&\csdef\cr
\kappa_s(\as(t),\as)&=\as(t) \kappa^\prime(\as)+\bar \kappa(\as)&\ksdef\cr
}
$$
are the linearized forms of $c$ and $\kappa$.
Solving eq.~\relosad\ for $M_s(N)$ gives
\eqn\losad{M_s(N)\equiv\half 
-\sqrt{\frac{\left(N-c(\as(t),\as)\right)}{\half\kappa(\as(t),\as)}},}
which, if one lets $\as(t)=\as$, reduces to the
standard naive dual 
\eqn\losadad{\gamma_s^B(\as,N)\equiv\half 
-\sqrt{\frac{\left(N-c(\as)\right)}{\half\kappa(\as)}},}
of $\chi_\sigma^q(\as,N)$
eq.~\finquad\ as it should.

Further terms in the expansion about the saddle generate an expansion
of the anomalous dimension in powers of
$\frac{N\beta_0}{d\varphi(N,M_s)/dM}$. In particular,
the fluctuations about the saddle lead to a contribution to the
anomalous dimension  which can be computed as in ref.~\sxrun:
\eqn\fluct{
\frac{1}{\sqrt{\phi'\left(N,M_s(\as(t),\as,N)\right)}}
=\exp\left[\int_{t_0}^t\!dt'\,\Delta \gamma_{ss}(\as(t'),\as,N)
\right] \frac{1}{\sqrt{\phi'\left(N,M_s(\as(t),\as,N)\right)}},}
where the prime denotes differentiation with respect to $M$ (which is
then set equal to $M_s$, eq.~\losadad), 
and
\eqn\fluctt{
\Delta \gamma_{ss}\left(\as(t),\as,N\right)\equiv -\half\beta_0 \as(t)
\frac{\varphi''(N,M) 
\varphi(N,M)}
{\varphi'^2(N,M)}
\Bigg|_{M=M_s}.}
Use of the saddle condition eq.~\losad\ with the $\varphi$ kernel
eq.~\phidef\ in eq.~\fluct\ gives, after setting $\as(t)=\as$,
\eqn\ssbat{
\gamma^B_{ss}\left(\as,N\right)=\gamma^B_{ss,0}
\left(\as,N\right)
+\smallfrac{1}{4}\as^2\beta_0\frac{c^\prime(\as)}
{\left(
c(\as)-N\right)},}
where
\eqn\gambzdef{\gamma^B_{ss,0}\left(\as,N\right)=-\beta_0\as +
\smallfrac{3}{4}\as^2\beta_0\frac{\kappa^\prime(\as)}{\kappa(\as)}.}
Note that $\gamma_{ss}^B$ depends on
both $\as/N$ and $\as$: in fact, it differs by terms of  order
$O(\as)$ from a contribution of order $\gamma_{ss}$ in the expansion
eq.~\sxexp\ of $\gamma_B$ in powers of $\as/N$ at fixed
$N$. The contribution
$\gamma^B_{ss,0}\left(\as,N\right)=\lim_{N\to\infty}\gamma^B_{ss}
\left(\as,N\right)$ 
is subleading in the expansion
eq.~\sxexp\ but leading--order  in the expansion eq.~\gammadef\ of the
anomalous dimension in powers of $\as$ at fixed $N$. It follows that, at
LO in the symmetrised DL expansion, as $N\to\infty$
\eqn\BatlargeN{\gamma_B(\as,N)\sim \gamma_s^B(\as,N)
+\gamma_{ss,0}^B(\as,N) + O(1/N).}

\topinsert
\vbox{
\epsfxsize=10truecm
\centerline{\epsfbox{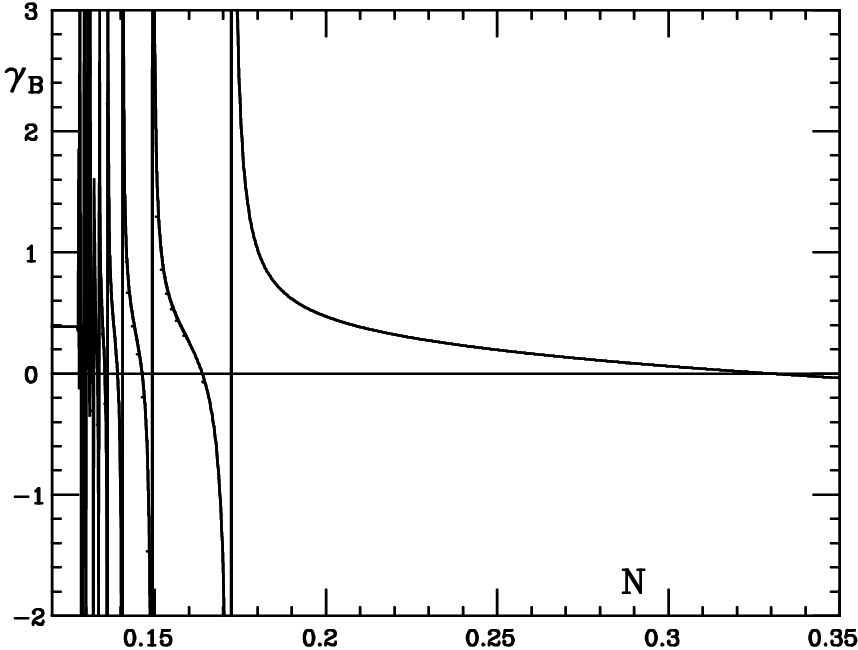}}
\bigskip
\hbox{
\vbox{\footnotefont\baselineskip6pt\narrower\noindent Figure 4: 
The Bateman anomalous dimension $\gamma_B(\as,N)$, computed  with
$\as=0.2$ and the values of the parameters which correspond to the LO
resummed curve of fig.~1, namely  $c=0.27$, $c^\prime=0.71$, $\bar
c=0.13$, $\kappa=1.40$, $\kappa^\prime=0.99$, $\bar\kappa=1.20$.}}
\hskip1truecm}
\endinsert

The Bateman anomalous dimension eq.~\batandim\ 
is displayed in fig.~4, with the values of the parameters which
correspond to the quadratic approximation to the LO resummed kernel
displayed in figs.~1,3. It is apparent that its behaviour is similar to 
that of the Airy  anomalous dimension discussed in ref.~\sxrun: the anomalous
dimension is regular and monotonically decreasing
along the real axis, with a simple pole at $N=N_B(\as)$, located at
the rightmost zero of the Bateman function $K_\nu(x)$. 
The small $x$ behaviour is
determined by the location of this pole, which in turn depends on the
values of the intercept and curvature of the quadratic kernel $c(\as)$
and $\kappa(\as)$ eq.~\finquad\ and their first derivatives 
$c'(\as)$, $\kappa'(\as)$.

The qualitative behaviour of the Bateman running coupling resummation
can be inferred from fig.~5, where we compare the quadratic
approximation to the LO resummed kernel (same as in fig.~3) with the
naive dual eq.~\dualdef\ of the physical branch of the Bateman
anomalous dimension (i.e. with $N>N_B(\as)$). For comparison, the
corresponding Airy anomalous dimension is also shown. The latter is
only defined when $c$ and $\kappa$ are linear functions of $\as$, and
it is thus determined by taking  in eq.~\finquad $c(\ahat)\approx
\ahat c(\as)/\as$ and $\kappa(\ahat)\approx \ahat
\kappa(\as)/\as$. This is a worse approximation than that used to
determined the Bateman anomalous dimension, since the leading log
terms are not correctly reproduced. 
The  resummation of running coupling effects effectively takes  
taking the minimum of $\chi$ from one half out to infinity~\sxrun, thereby 
greatly extending the range of the physical region. For the anomalous 
dimension this means that the usual fixed coupling cut at 
$N=c(\as)$ is replaced by a simple pole at $N_B(\as)$ to the left, i.e. 
$N_B(\as)<c(\as)$. This in turn means that the small $x$
behaviour of the splitting function will be softer 
(more slowly rising)
than that expected from naive duality.  

In fig.~5 (right) we display
the Bateman effective kernel 
determined after subtracting from the
Bateman anomalous dimension its $N\to\infty$ limit, namely, the dual 
of $\gamma_B-\gamma^B_{ss,0}$, with
$\gamma^B_{ss,0}$ given by eq.~\gambzdef, together with the corresponding Airy
curve. This comparison shows that, after this constant subtraction,
for $M\leq 0$ $\gamma_B$  is very well approximated by the
 leading order asymptotic $\beta\to0$ approximation eq.~\losadad,
 i.e., by the  fixed-coupling quadratic curve, as is
 $\gamma_A$. In other words, the running coupling resummation for
 $N>c(\as)$ (to the right of the fixed--coupling cut) is a very small 
correction.

\topinsert
\vbox{
\epsfxsize=8truecm
\centerline{\epsfbox{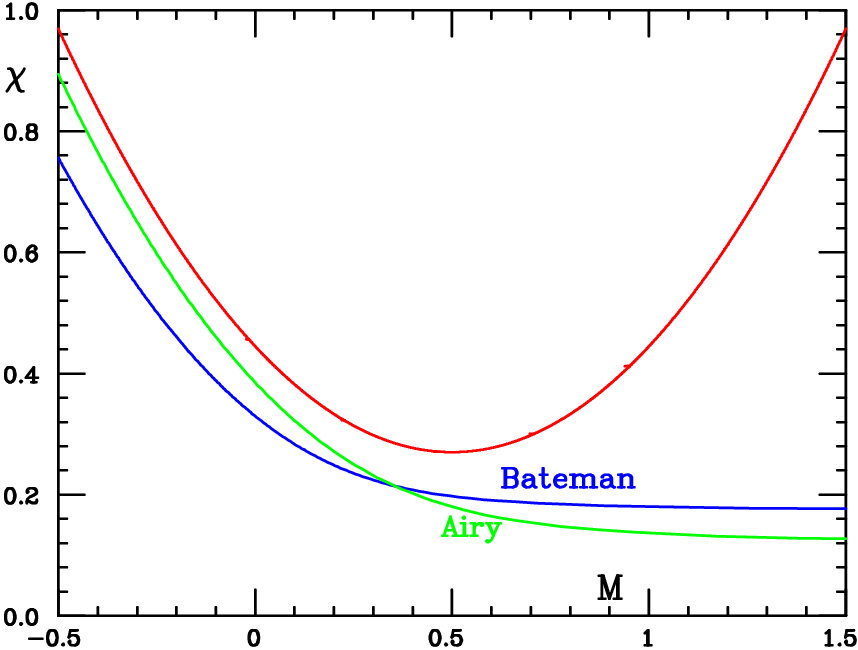}\epsfxsize=8truecm\epsfbox{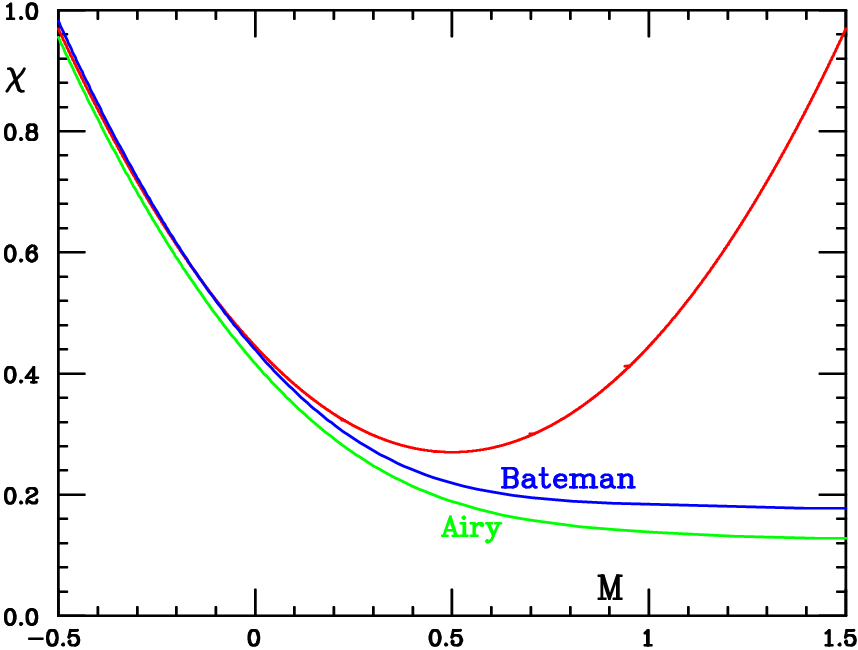}}
\bigskip
\hbox{
\vbox{\footnotefont\baselineskip6pt\narrower\noindent Figure 5: (left)
The
quadratic kernel (upper curve) obtained from the expansion near the
minimum of the LO resummed curve in fig.~3, compared with the
corresponding effective Bateman kernel. The effective Airy kernel,
also shown for comparison, is determined  with the same value of $c$
and $\kappa$, i.e. approximating $c(\ahat)\approx \ahat c(\as)/\as$
and  $\kappa(\ahat)\approx \ahat \kappa(\as)/\as$. \hfill\break
 (right) Same as in the left figure, but with the effective kernels 
determined from the Bateman and
 Airy anomalous dimension after subtraction of the asymptotic large
 $N$ contribution $\gamma_{ss,0}^B$. 
}}\hskip1truecm} 
\endinsert

The location of the rightmost singularity of the anomalous dimension,
which determines the small $x$ behaviour, is shown in fig.~6 as a
function of $\as$. 
This singularity is a cut at fixed coupling, and a
simple pole after running coupling resummation. The nonlinear
dependence of $c$ on $\as$ in the resummed fixed-coupling case (as
opposed to the linear dependence in the unresummed BFKL LO case) is
apparent, and it explains why a linear extrapolation from 
the origin would not be a good approximation for realistic values of $\as$.

\topinsert
\vbox{
\epsfxsize=12truecm
\centerline{\epsfbox{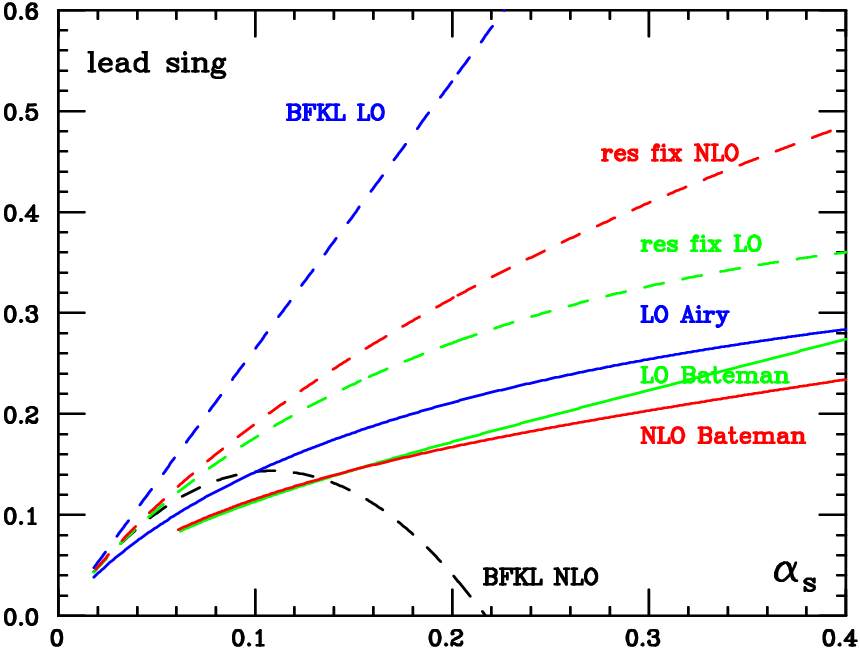}}
\bigskip
\hbox{
\vbox{\footnotefont\baselineskip6pt\narrower\noindent Figure 6: The
leading (rightmost) singularity of various anomalous dimension as a
function of $\as$, all determined with $n_f=0$. 
BFKL LO , res fix LO and res fix NLO (dashed) denote
respectively the
location of the cut for $\as\chi_0$ eq.~\kizero\ and for the
$\beta_0=0$ resummed kernels $\chi_{\sigma,\, LO}$ and
$\chi_{\sigma,\, NLO}$
obtained putting on shell the kernels eq.~\sigLODIS\ and \sigNLODIS.
LO Airy, LO Bateman and NLO Bateman (solid curves) denote the 
location of the pole
$N_B$ for the running coupling resummation of the corresponding three curves.
BFKL NLO (dashed) denotes the location of the stationary point of the NLO
kernel eq.~\chiexp.
}}\hskip1truecm} 
\endinsert

In all cases, running
coupling resummation softens the leading singularity. The effect is
most dramatic for the LO BFKL kernel $\ahat \chi_0$ eq.~\kizero, whose
running coupling resummation is effected by the Airy method of
ref.~\sxrun. Even though the percentage softening is more moderate for
the Bateman resummation discussed here (see also fig.~5), the symmetrization 
discussed
in the previous section also reduces the fixed coupling intercept in
comparison to the BFKL one. Hence, after running coupling resummation
the LO linear (Airy) result and the NLO resummed (Bateman) result for
the location of the pole end up being quite close. The stability of
the resummed result when going from LO to NLO is further enhanced by
running coupling resummation, so all three running coupling curves end up
being close to each other. Interestingly, the contributions to the NLO kernel
proportional to $\beta_0$ (to be discussed in detail in section~6)
further reduce the difference between the resummed LO
and NLO minima, and in fact they pull the NLO resul below the LO one
for $\as\gsim 0.15$.
For comparison, we also show the location
of the NLO unresummed BFKL stationary point, though this is in fact 
unphysical for
realistic $\as\gsim 0.1$ where the NLO BFKL kernel in fact has a
minimum instead of a maximum, so the associated anomalous dimension
becomes imaginary and the cross section oscillates~\refs{\brus,\ross}.

The Bateman anomalous dimension will be used in the next sections to
construct the improved anomalous dimensions in the symmetrised 
double leading expansion, first at LO and then at NLO.

\newsec{Leading Order Resummation}

We wish now to determine anomalous dimensions and splitting functions
by combining the ingredients of the previous two sections:
symmetrization of the double--leading expansion and its running
coupling resummation. In previous work~\refs{\sxrun,\sxphen}\ we 
constructed resummed anomalous dimensions by exploiting the fact that
the DL expansions of $\chi$ and $\gamma$ are dual of each other, up to
subleading terms. We thus constructed the anomalous dimension
$\gamma$ starting from its own DL expansion, namely
\eqn\dlgam{\eqalign{
\gamma_{DL}(\as,N) &= \gamma_{DL\,LO}(\as,N)
+\as \gamma_{DL\,NLO}(\as,N) +\dots\cr
\gamma_{DL\,LO}(\as,N)&= [\as\gamma_0(N)+\gamma_s(\smallfrac{\as}{N})
-\smallfrac{n_c\as}{\pi N}]\cr
\gamma_{DL\,NLO}(\as,N)&= [\as\gamma_1(N)+\gamma_{ss}(\smallfrac{\as}{N})
-e_0-\as\left(\frac{e_2}{N^2}+\frac{e_1}{N}\right)].}}
Here, instead, in order to take advantage of the improved kernel
 $\chi_{\Sigma\,LO}(\as,M)$, 
which is the on-shell version of eq.~\sigLODIS, we must define the
 anomalous dimension $\gamma_{\Sigma\,LO}(\as,N)$ 
using the duality equation \expsimn: 
\eqn\siggam{\chi_{\Sigma\,LO}(\as,\gamma_{\Sigma\,LO}(\as,N))=N.}
This coincides with the LO double--leading 
anomalous dimension $\gamma_{DL\,LO}$ up to subleading terms.   

To go beyond the fixed coupling limit, in the kernel $\chi_\Sigma(\as,M)$  
we must make the replacement $\as\to\hat\as$. As proven in Ref.~\sxrun,
factorized duality remains valid order by order in perturbation theory
at the running coupling level,
i.e. for any given $\chi$  there exists a $\gamma$ such
that the solution to eq.~\asheq\ with the given $\chi$ and the
solution to the usual evolution equation~\tevol\ coincide if, order by
order, the 
boundary conditions are suitably matched and appropriate
$\beta_0$-dependent corrections are added to the fixed coupling
kernels. In particular, in the LO expression of  $\chi_\Sigma(\as,M)$
eq.~\sigLODIS, the terms originating from $\chi_0$ are not modified
and, given the basic commutator relation, eq.~\bascomm, they do not
depend on operator ordering up to NLO corrections. For the  $\chi_s$
terms it is easy to see that an 
operator ordering exists such that the $\chi_s$ which is dual to $\gamma_0$
coincides
with the fixed--coupling expression eq.~\kis. 
Indeed, the leading--order form of the duality relation eq.~\dualinv\
in the running coupling case is
\eqn\lorundual{\gamma_0(\chi_s)=\hat\as^{-1} M,}
which immediately implies that if $\chi_s$ satisfies eq.~\kis, i.e. it
is the inverse function of $\gamma_0$, then 
\eqn\rcdual{
\chi_s\left[\gamma_0\left((\ahat^{-1}
M)^{-1}\right)\right]=\chi_s\left(M^{-1}\ahat\right)=N}
by direct substitution of
eq.~\lorundual. Hence, the running-coupling dual of $\gamma_0$
coincides with the fixed coupling dual $\chi_s$, provided $\chi_s$ is
now evaluated as a function of the operator $M^{-1}\ahat$. A more
detailed discussion of operator ordering issues in running coupling
duality will be given elswehere~\comm. 

Let us now turn to the symmetrization of $\chi_s$. First, we should
observe that the symmetry of the underlying Feynman diagrams which
determines the symmetry of the BFKL kernel $K\left(\alpha_s,\smallfrac{Q^2}{k^2}\right)$ 
eq.~\bfklker\ holds for the unintegrated gluon distribution ${\cal
G}(\xi,Q^2)$, obtained by differentiation from the standard gluon distribution 
which enters the GLAP equation:
\eqn\unintdef{ {\cal
G}(\xi,Q^2)\equiv\frac{d}{dt} G(\xi,Q^2),}
i.e., upon Mellin transformation,
\eqn\unintdefm{ {\cal
G}(\xi,M)\equiv M G(\xi,M).}
When the coupling runs, the $M$--space evolution kernel acquires a
further contribution when switching from the integrated to
unintegrated gluon distribution due to the commutator of $\ahat$ and
the $M$ prefactor in eq.~\unintdefm. Indeed, 
\eqn\commiuis{\eqalign{ \chi_s \left(M^{-1}\ahat\right) G(\xi,M) &=\chi_s
\left(M^{-1}\ahat\right) M^{-1} {\cal
G}(\xi,M) \cr &= M^{-1} \chi_s \left(\ahat M^{-1} \right){\cal
G}(\xi,M).}}
So, going from the integrated to the unintegrated distribution switches the
ordering of $\ahat$ and $M^{-1}$ in the argument of $\chi_s$. Note
that this transformation is manifestly not symmetric upon $M\to 1-M$,
hence if the kernel at the unintegrated level is symmetric, the kernel
at the integrated level won't be.
 
At the running
coupling level, the kernel is symmetric 
only if the argument of the running coupling is also treated
symmetrically. In fact, upon Mellin transformation, a
 $\xi$ evolution equation of the form
\eqn\xevolrunqlr{\frac{\partial}{\partial\xi}G(\xi,Q^2)=\int_{-\infty}^{\infty}
\frac{dk^2}{k^2}  \sum_{p=1}^\infty
\left[\as^p(Q^2) K_L^{(p)}(Q^2/k^2) +  \as^p(k^2)
K^{(p)}_R(Q^2/k^2)\right] G(\xi,k^2)}
becomes
\eqn\xevolrunmlr{\frac{\partial}{\partial\xi}G(\xi,M)=
\sum_{p=1}^\infty\left[ \ahat^p \chi_L^{(p)}(M)  +  
\chi^{(p)}_R(M)\ahat^p\right] G(\xi,M),}
where $\chi^{(p)}_L(M)$,  $\chi^{(p)}_R(M)$ are the Mellin transforms
eq.~\Mmom\ of the kernels $K^{(p)}_R(Q^2/k^2)$ and
$K^{(p)}_L(Q^2/k^2)$, respectively. Because the symmetrization is
based on the interchange $Q^2\leftrightarrow k^2$, it follows that the
symmetrized version of
$\ahat^k f(M) $  is $\half\left(\ahat^k f(M) + f(1-M) \ahat^k\right) $. 
Furthermore, because
\eqn\comm{[\ahat , M^{-1}]=- \beta_0 M^{-1}\ahat^2 M^{-1}=- \beta_0
\ahat M^{-2}\ahat,}
while
\eqn\recomm
{[\hat \as, (1-M)^{-1}]=\beta_0 (1-M)^{-1}\hat\as^2 (1-M)^{-1}
=-\beta_0\hat\as (1-M)^{-2}\hat\as,} 
it is easy to show that if 
$$f( \ahat M^{-1} )=\tilde f(M^{-1} \ahat  ) =\sum_k
\tilde f_k  \ahat^k M^{-k} $$ then  $$f( (1-M)^{-1}\ahat )=\tilde f(
\ahat (1-M)^{-1} ) =\sum_k 
\tilde f_k   (1-M)^{-k} \ahat^k. $$

It follows that at the running-coupling level the symmetrized LO kernel
is \eqn\simmLORC{
\bar\chi_{\sigma\,LO}(\ahat,M,N)
=\chi_s\left(\ahat \left(M+\smallfrac{N}{2}\right)^{-1}\right)+ 
\chi_s\left(\left(1-M+\smallfrac{N}{2}\right)^{-1}\ahat \right)  
+\tilde\chi_{0}(\ahat,M,N)+\chi_{\rm mom},}
where
\eqn\coldecrc{\eqalign{\tilde\chi_{0}(\ahat,M,N)&=
\ahat\Big(\bar\chi^+_0(M,N)-\smallfrac{1}{M+\smallfrac{N}{2}}\Big)
+\Big(\bar\chi^-_0(M,N)-\smallfrac{1}{1-M+\smallfrac{N}{2}}\Big) \ahat\cr
&=\smallfrac{n_c}{\pi}\left[\ahat\left(\psi(1)
-\psi(1+M+\smallfrac{N}{2})\right)+
\left(\psi(1+N)
-\psi(2-M+\smallfrac{N}{2})
\right) \ahat
\right]
,}} 
which are the running coupling analogues of
eqs.~\simmLO,~\chicollos.
Different orderings of the $\chi_s$ term
can be determined using the commutators \comm\ and \recomm.  
Different orderings of the remaining terms
are  equivalent, up to subleading corrections which modify the form of
the subleading $\chi_1$, and will be discussed in the next
section. Here, we have chosen the operator ordering in the $\chi_0$ terms such
that it matches the ordering of the corresponding $\chi_s$ terms,
namely, so that the ordering is the same in the double counting
contributions. Note that this corresponds to taking, after inversion
of the Mellin transform, $\as(Q^2)$ in the collinear
double--counting term, and  $\as(k^2)$ in its anti-collinear 
counterpart. 

Starting from eq.~\simmLORC~ one can expand
$\bar\chi_{\sigma\,LO}(\ahat,M,N)$ in powers of $(M-\half)$ at fixed
$N$ and $\ahat$, around its minimum at $M=\half$ with a result of the
form: 
\eqn\rcreshuf{\bar\chi_{\sigma}(\ahat,M,N)=\bar\chi^q_{0}(\ahat,M,N)
+\bar\chi^q_{s}(\ahat,M,N)+O(\beta_0\ahat)+O((M-\half)^4),}
where 
\eqn\chizq{
\bar\chi^q_{0}(\ahat,M,N)=\ahat\left(c_0(N)
+\half\kappa_0(N)(M-\half)^2\right),}
while
\eqn\chisqdef{\bar\chi^q_{s}(\ahat,M,N)
= \bar c(\ahat,N) 
+\half \bar\kappa(\ahat,N) \left(M-\half
\right)^2,}
and the expansion coefficients are given by
\eqn\rccoeff{\eqalign{
\bar c(\ahat,N)&=2c_s\left(\smallfrac{\ahat}{1+N}\right),\cr
\bar \kappa(\ahat,N)&=2\smallfrac{1}{(1+N)^2}\kappa_s
\left(\smallfrac{\ahat}{1+N}\right),\cr
}}
in terms of the coefficients of the Taylor expansions of 
$\chi_s$: 
\eqn\chisexp{\chi_s\left(\smallfrac{\as}{M}\right
)= c_s(\as) + \delta_s(\as) \left(M-\half \right) 
+\half \kappa_s(\as) \left(M-\half \right)^2+\dots}
The on-shell quadratic expansion eq.~\finquad\ can now be obtained by 
putting $N$ on shell, i.e. by solving $N = \bar\chi_{\sigma}(\ahat,M,N)$
with the right-hand side given by eq.~\rcreshuf\ for 
$N$ consistently to order $(M-\half)^2$, and then identifying the
result $N= \chi_\sigma(\ahat,M)$ . 
Alternatively one of course could first obtain the full on-shell 
$\chi_\sigma(\ahat,M)$ by solving eq.~\expsimm\ and then expanding the 
result about $M=\half$: the result will be the same.

The $O(\beta_0\ahat)$ corrections in eq.~\rcreshuf\ are due to operator ordering, and will be
discussed in sect.~6. They are of relative order $\ahat$ with respect
to the curvature and intercept, which are determined to all orders in
$\ahat$ in order to obtain a good description of the $\chi$ kernel in
the vicinity of the minimum. Further $O(\beta_0\ahat)$ corrections,
also to be discussed in the next section, are
generated when switching from the unintegrated distribution (which
evolves with a symmetric kernel) to the standard integrated one which
enters the GLAP equation. 

The final expression for the improved anomalous dimension at LO is now
obtained by implementing the running coupling resummation through the
quadratic approximation. Namely, we
write the evolution equation eq.~\rcee\ as
\eqn\symnevolrun{
NG(N,M)=\chi_\sigma(\ash,M) G(N,M)+F(M),}
and then exploit the symmetry of $\chi_\sigma$ to expand it about its
minimum, up to a quartic term which leads to
asymptotically subleading~\refs{\sxrun,\runph} corrections. The
anomalous dimension determined from solution of eq.~\symnevolrun\ in the
quadratic approximation is then simply given by $\gamma_B$
eq.~\batandim, but with the replacement $\half\to \half+\half N$. The anomalous
dimension in asymmetric (DIS) variables is obtained from it using
eq.~\gsymnonsym, which simply restores $M_s$ to the value eq.~\batandim.
 
Hence, the resummed anomalous dimension after running coupling
resummation is given by
\eqn\rcres{\eqalign{&\gamma_{\Sigma\,LO}^{rc}(N,\as(t))=
\gamma_{\Sigma\,LO}(N,\as(t))+\gamma^B(\as(t),N)\cr&\qquad- 
\gamma^B_s(\as(t),N)-\gamma^B_0(\as(t),N)+\gamma_{\rm mom}(\as(t),N),\cr}}
where $\gamma_{\Sigma\,LO}(N,\as(t))$ is determined from eq.~\siggam,
$\gamma^B(\as(t),N)$ is the Bateman anomalous dimension
eqs.~\batandim, and  $\gamma^B_s(\as(t),N)$, 
$\gamma^B_0(\as(t),N)$, obtained letting $\as \to \as(t)$ in
eqs.~\losadad,\gambzdef, remove the double counting between the 
the first two  terms. Notice that $\gamma^B_s$ removes the
square--root branch cut of the fixed-coupling anomalous dimension,
thanks to the fact that the curvature and intercept of the quadratic
kernel are the same eq.~\ckrel\ in
symmetric variables (used to compute $\gamma_B$) and DIS variables
(used to compute $\gamma_\Sigma$). 
The momentum subtraction term is needed to cancel small (and formally 
subleading) momentum violations incurred during the running coupling 
resummation: it can be chosen as
\eqn\momsubnew{\gamma_{\rm mom}(\as(t),N)=d_m f_m(N),}
where $f_m$ must satisfy $f_m(1)=1$ and $f_m(\infty)=0$ \runph\ and
can thus be implemented through the same function $f_m$ introduced in
eq.~\fdef, and $d_m$ is a constant of order $\as^2$.  

\newsec{Next-to-Leading Order Resummation}
The extension of our results to next-to-leading order requires two
steps: first, the symmetrization of the next-to-leading DL kernel
eq.~\kiDLNLO, and second, the next-to-leading treatment of the running
coupling resummation.
The general structure of the next-to-leading order symmetrized
kernel  was already given in
eqs.~\simmNLO,\sigNLODIS\ in the fixed coupling case. At the running
coupling level, operator ordering in the leading order contribution
affects the form of the NLO term.
We assume therefore that (in symmetric variables)
\eqn\kisym{\bar\chi_{\sigma}(\ahat,M,N)=
\bar\chi_{\sigma\,LO}(\ahat,M,N)+\ahat
\bar\chi_{\sigma\,NLO}(\ahat,M,N),}
where $\bar\chi_{\sigma\,LO}(\ahat,M,N)$ is given by
eq.~\simmLORC. 
The corresponding kernel in DIS variables is
\eqn\kiasym{\bar\chi_{\Sigma}(\ahat,M,N)=\bar\chi_{\sigma}
(\ahat,M-\smallfrac{N}{2},N).}
  
 Note that at the running coupling level the kernel for
evolution of
the integrated and unintegrated parton  distributions are not equal,
as shown in eq.~\commiuis\ for $\chi_s$, while in fact, as discussed
in the previous 
section, only 
the unintegrated distribution is symmetric. Hence, we will construct
the symmetrized $\chi$ kernel at the
unintegrated level, and then switch back to the integrated level at
the end when determining the resummed GLAP anomalous dimension from it.

At the leading order level, 
we have already shown in section~3 that the fixed coupling
$\bar\chi_{\Sigma\,LO}(\as,M,N)
=\bar\chi_{\sigma\,LO}(\as,M-\smallfrac{N}{2},N)$ 
is equal to $\chi_{DL\,LO}(M,N)$ up to
subleading terms, and generalized this in section~4 to the running
coupling $\bar\chi_{\Sigma\,LO}(\ahat,M,N)$. We must now check this
equality up to NLO at the running coupling level, and compute the
correction terms which are necessary to restore it when violated. This
will lead to the determination of $\bar\chi_{\Sigma\,NLO}(\ahat,M,N)$
eq.~\kiasym, and in particular of the function
$\bar \chi_1(M)$ which was left implicit in  eqs.~\simmNLO,\sigNLODIS.

We start therefore with  $\bar\chi_{\Sigma\,LO}(\ahat,M,N)$ 
given by eq.~\simmLORC. 
 This differs from the  double leading $\chi_{DL\,LO}(M,N)$ in two
respects: because of the addition of the anticollinear
term $\chi_{s}( (1-M+N)^{-1}\ahat)$, and because 
$\chi_0$ is put off shell.
In order to compare to the DL result it is convenient
to view the simple and double poles $(1-M+N)^{-1} \ahat $ and
$( (1-M+N)^{-1} \ahat)^2$ of the
anticollinear term as contributions to the off-shell $\chi_0$ and
$\chi_1$, respectively.

Consider first the anticollinear term  $\chi_s( (1-M+N)^{-1} \ahat)$ of 
eq.~\simmLORC. 
This   term starts at $O(\as)$, but the first two terms are already
included in $\chi_0$ and $\chi_1$, hence the remaining contribution
starts at $O(\as^3)$, namely NNLO in an expansion in powers  of $\as$
at fixed
$M$, and N$^4$LO in an expansion in powers  of $\as$
at fixed $\as/M$.
Let us turn now to the rather more complicated case of the 
off-shell extension  of $\chi_0(M)$, which is given by
the $O(\ahat)$ contribution
to  $\bar\chi_{\Sigma\,LO}(\ahat,M,N)$  (in an expansion in
powers of $\ahat$ at fixed $M$). It is thus  
\eqnn\locont\eqnn\loreorder$$\eqalignno{&
\ahat \bar\chi_0^+(M-\smallfrac{N}{2},N)
+ \bar \chi_0^-(M-\smallfrac{N}{2},N)\ahat &\locont\cr
&\qquad\quad
=\ahat \left[ \bar\chi_0^+(M-\smallfrac{N}{2},N)
+\bar \chi_0^-(M-\smallfrac{N}{2},N) 
\right]-
\ahat^2 \beta_0 \smallfrac{n_c}{\pi}\psi^\prime(1-M)+O(\ahat^3) ,&\loreorder\cr}$$
where $\bar\chi_0^\pm(M,N)$ are given by eq.~\chicollos, and
in the last step we 
used the commutator eq.~\bascomm, which implies
\eqn\funcomm{[\ahat,f(M)]=\ahat^2\beta_0f^\prime(M)
-\ahat^3\beta_0^2f^{\prime\prime}(M)+\dots}
The ordering of eq.~\loreorder\ for the $O(\as)$ contribution in the
expansion eq.~\chiexp\ of the kernel has been adopted in
refs.~\refs{\sxres{--}\sxphen\sxrun\runph} for the construction of the
double--leading expansion, though the  ordering eq.~\locont, as
discussed in sect.~4, is more convenient in a symmetrized approach.

Let us first consider the effect of the off-shell extension
on terms which are next-to-leading in $\ln Q^2$, i.e. in 
the expansion of $\chi$ in powers of $\as$ at fixed
$\as/M$.  The effect of the off-shell extension can be determined 
by considering
\eqn\chiexpllq{\eqalign{&\left[\chi_0^+(M)+
\chi_0^-(M-N)\right]-\chi_0(M)
=\smallfrac{n_c}{\pi}\left[\psi(1-M)-\psi(1-M+N)\right]\cr
&\qquad\qquad=\smallfrac{n_c}{\pi}
\left\{\left[\psi(1)-\psi(1+N)\right]-M
\left[\psi^\prime(1)-\psi^\prime(1+N)\right]\right\}+O(M^2),}} 
where $\chi^\pm(M)$ are given by eq.~\chicoll.
It follows that
\eqn\chiexpllqs{\ahat \bar\chi_0^+(M-\smallfrac{N}{2},N)
+ \bar\chi_0^-(M-\smallfrac{N}{2},N)\ahat 
-\ahat\chi_0(M)=O(M)+O(\ahat^2),}
where now $\bar\chi_0^\pm(M,N)$ is given by eq.~\chicollos, and
the $O(\ahat^2)$ correction eq.~\loreorder\ due to operator ordering
will be included in $\chi_1$ as we discuss below.
Therefore, thanks to the replacement
$\psi(1)\to\psi(1+N)$ in $\bar\chi^-(M,N)$ eq.~\chicollos\ in comparison to $\chi^-(M)$
eq.~\chicoll,
the combination eq.~\loreorder\ which appears in the leading--order kernel
$\bar\chi_{\Sigma\,LO}$, eq.~\sigLODIS, differs from the BFKL
kernel $\as\chi_0$, eq.~\kizero, by terms which are of order $O(\as M)=
O(\as^2(\smallfrac{\as}{M})^{-1})$ i.e. next-to-next-to leading 
$\ln Q^2$ in
an expansion in powers of $\as$ at fixed
$\as/M$. 

Next we consider the effect of the off-shell extension on terms which
are next-to-leading $\ln x$, i.e. the expansion  of the kernel in 
powers of $\as$ at fixed
$N$. Because this expansion of the kernel (and in particular the NLO
term $\chi_1(M)$) is usually computed in symmetric variables, it is useful
to compare the combination which appears in the leading--order kernel
$\bar\chi_{\Sigma\,LO}$, eq.~\sigLODIS\ with  the LO BFKL
kernel $\as\chi_0$ in symmetric variables, too.
It turns out that the off-shell extension of $\chi_0$ 
induces next-to-leading $\ln
\smallfrac{1}{x}$ contributions which must be subtracted. Indeed,
\eqn\chiexpllx{[\bar\chi^+_0(M,N)+\bar\chi^-_0(M,N)]-\chi_0(M)
=\half\smallfrac{n_c}{\pi}\left[2\psi'(1)-\psi^\prime(M)-
\psi^\prime(1-M)\right]N+O(N^2).
}
In order to construct the appropriate subtraction,
 start with the NLO BFKL kernel eq.~\chiexp, written in
symmetric variables. The NLO version of eq.~\locont\ is (in 
symmetric variables)
\eqn\chibfklnlo
{\chi_{\sigma\,NLO}(\ahat,M,N)=\ahat\chi_0^+(M,N)+
 \chi_0^-(M,N)\ahat+\ahat^2\chi_1(M,N)+O(\ahat^3),}
where
the ordering in the NLO term $\chi_1(M)$ is immaterial since it only
affects NNLO terms. 

With the symmetric ordering of $\ahat$ adopted in
 eq.~\loreorder, the NLO contribution
$\chi_1(M,N)$ 
is symmetric. The on-shell contribution $\chi_1(M)$ has been determined 
in refs.~\refs{\fl\fleta\dd{--}\cc}, in the form
\eqnn\flresult\eqnn\flresultreord$$\eqalignno{\chi(\ahat,M)&=
\ahat\chi_0(M)+\ahat^2 \chi_1^{FL}(M)+O(\ahat^3)&\flresult\cr
&= \ahat\chi_0^+(M) +
\chi_0^-(M)\ahat +\ahat^2\left[\chi_1^{FL}(M)+\beta_0
\smallfrac{n_c}{\pi}\psi^\prime(1-M)\right]
+O(\ahat^3),&\flresultreord}$$
where $\chi_1^{FL}(M)$  is given
by $\delta(\gamma)$ in ref.~\fl, with the identification $\gamma\to M$
and $\chi_1^{FL} = \frac{n_c^2}{4\pi^2}\delta$. In practice, the
effect of the $\beta_0$ term in eq.~\flresultreord\ is thus to reverse 
the sign of the
$\psi^\prime(1-M)$ contribution to $\Delta(\gamma)$ (as defined in ref.~\fl)
thereby leading to a manifestly symmetric result for 
\eqn\symflchidef{\chi_1(M)\equiv 
\chi_1^{FL}(M)
+\beta_0\smallfrac{n_c}{\pi}\psi^\prime(1-M).}
We can now finally remove the terms generated by the off-shell extension
eq.~\chiexpllx\ up to NLO by defining (in asymmetric variables)
\eqnn\chinllx\eqnn\chitil$$\eqalignno{&\chi_{NLLx}(\ahat,M)=\ahat\chi_0^+(M)
+\chi_0^-(M-N)\ahat +\ahat^2 \breve \chi_1(M)+ O(\ahat^3)&\chinllx\cr
&\quad\breve\chi_1(M)=\chi_1(M)-\half\chi_0(M)\smallfrac{n_c}{\pi}
\left[2\psi'(1)-\psi^\prime(M)-
\psi^\prime(1-M)\right].&\chitil}$$
With this definition, the off-shell extension eq.~\chinllx\ coincides
with the on-shell $O(\ahat^2)$ kernel of
refs.~\refs{\fl\fleta\dd{--}\cc}, up to sub-subleading corrections.

Note that although $\chi_1(M)$ has triple poles at $M=0$ and $M=1$, 
these are cancelled in $\breve\chi_1(M)$, so that the leading 
small $M$ singularity is $O\left(\smallfrac{\as^2}{M^2}\right)$ as 
required by the matching to $\chi_s$. This cancellation would be 
spoiled if one did not replace $\chi_1$ with $\breve\chi_1$ on 
the r.h.s. of eq.~\chitil. Of course this is all as it should be:
we obtained $\breve\chi_1$ through eq.~\chiexpllx\ as compensation for 
the off-shell extension of $\chi_0$ eq.~\chicollos, which was in itself
constructed precisely so that LO collinear evolution is not spoiled
at large $N$. The cancellation of the triple poles is then the NLLx 
manifestation of this fact.

Therefore, to match to the collinear contributions $\chi_s$ and 
$\chi_{ss}$, and their anticollinear counterparts, as described by
eq.~\simmLO\ and eq.~\simmNLO, we define a function $\tilde\chi(M)$ as
\eqn\chibaronedef{
\tilde \chi_1(M)= \breve\chi_1(M)-\left[g_1\left(\smallfrac{1}{M}+
\smallfrac{1}{1-M}\right)+g_2\left(\smallfrac{1}{M^2}+
\smallfrac{1}{(1-M)^2}\right)\right],}
with coefficients $g_i$ are determined by demanding regularity as $M\to0$ and 
$M\to1$, and thus correct matching to $\chi_s$ and $\chi_{ss}$:
\eqn\gcoeff{\eqalign{ 
g_1 &= -\smallfrac{n_f n_c}{\pi^2}\left(10+\smallfrac{13}{n_c^2}\right),\cr 
g_2 &= -\smallfrac{11n_c^2}{12\pi^2} 
- \smallfrac{3n_f}{4\pi^2 n_c}.\cr}}

Now, note that, of course, a shift by an amount proportional to $N$ of
the argument of $\chi_1$ is sub-subleading: $\tilde\chi_1(M+ k N)=
\tilde\chi_1(M)+O(\as)$. Hence, we can simply take in 
eqs.~\simmNLO,\sigNLODIS\
\eqn\onshellchione{\tilde\chi_1(M,N)=
\tilde\chi_1(M-\smallfrac{N}{2},N)+O(\as)=\tilde\chi_1(M),}
with $\tilde\chi_1(M)$ given by eq.~\chibaronedef. With this choice,
$\chi_\Sigma(M)$ obtained putting on shell $\bar\chi_\Sigma(M,N)$
coincides with its double--leading counterpart up to subleading
terms: eq.~\chiexpllqs\ shows that the off-shell extension of $\chi_0$
only adds  terms
which are subleading in an expansion in powers of $\as$ at fixed
$\as/M$,
and eq.~\chitil\ shows that it only adds  terms
which are subleading in an expansion in powers of $\as$ at fixed
$M$, while the anticollinear contribution is also subleading.

The choice of $\bar\chi_1$ eq.~\onshellchione\  has two shortcomings,
however. First,
eq.~\onshellchione\ implies that the pair of kernels in symmetric and
asymmetric variables are only related by the shift eq.~\symnonsym\ up
to subleading corrections as well, because the contribution $\bar \chi_1$ does
not satisfy eq.~\symnonsym. This in particular means that also the
equality of curvature and intercept eq.~\ckrel\ of the pair of kernels
is spoiled by subleading terms. As discussed in sect.~4, it  then
follows that the cancellation
of the square-root cut in eq.~\rcres\ is spoiled by subleading
corrections. Second, as discussed in sect.~3, $\chi_\Sigma$ in
symmetric variables is an
entire function, provided that eq.~\symnonsym\ holds. This property is
lost if eq.~\onshellchione\ is used, and $\chi_\Sigma$ then has
subleading poles on the real axis, so that in particular the quadratic
approximation is much poorer.

It is therefore necessary to construct $\chi_1^\pm(M,N)$ such that
while it is symmetric in symmetric variables, in asymmetric variables 
 $\chi_1^\pm(M-\smallfrac{N}{2},N)$ reproduces the correct LO and NLO 
singularities, as in eq.~\chibaronedef, even at large $N$, while at small 
$N$ $\bar\chi_1(M,N)=\tilde\chi_1(M)+O(N)$. The construction follows the 
same procedure as at LO, in particular by first separating
\eqn\offshellchione{\chi_1(M)=\chi^+_1(M)+\chi^-_1(M),} 
where $\chi^+_1(M,N)$ ($\chi^-_1(M,N)$) is regular when $\Re M>\half$ 
( $\Re M<\half$) just as in eq.~\chicollos. However the details are 
somewhat technical, and have been relegated to an appendix. 

Because operator ordering issues are only relevant in the leading
order term,\foot{Strictly speaking, at the running coupling level
the expression eq.~\kiss\ of $\chi_{ss}$ is 
corrected by $O(\beta_0)$ terms which
are of the same order as $\chi_{ss}$~\comm. However, since our final aim is
the determination of the anomalous dimension, we will neglect these
terms, which do not affect the final double-leading anomalous
dimension, but only the NLO symmetrization.}
this concludes the construction of $\chi_{\Sigma}$ eq.~\kiasym: the
NLO kernel $\chi_{\Sigma\,NLO}$
is found using the kernel $\bar\chi_1(M,N)$, which
reduces to $\chi_1(M)$ as $N\to0$ in eqs.~\simmNLO\
or~\sigNLODIS, while the collinear and anticollinear terms are given
by $\chi_{ss}$ eq.~\kiss, with argument 
$M^{-1}\ahat$ in the collinear and
$\ahat M^{-1}$ in the anticollinear terms, and off-shell
extension as in eqs.~\simmNLO,\sigNLODIS.

Let us now turn to the running-coupling resummation. First,
we must determine the quadratic approximation to the NLO kernel
obtained by putting on shell $\bar\chi_\Sigma$ eq.~\kisym. Operator
ordering issues are only relevant in the leading-order terms, as
otherwise they lead to sub-subleading corrections, whereas different
choices of operator ordering in $\bar\chi_{\Sigma\,LO}$ lead to terms
of order $\bar\chi_{\Sigma\,NLO}$. 

The leading--order kernel $\bar\chi_{\sigma\,LO}$
is determined with the canonical ordering eq.~\simmLORC, whereas the
quadratic kernel eq.~\finquad\ used in the running-coupling
BFKL equation~\rceeexp\ is based on the ordering eq.~\loreorder, with
all $\ahat$ to the left. The quadratic approximation thus receives two
contributions due to operator reordering. 
The first is due to reordering in $\bar\chi_0(M,N)$ eq.~\coldecrc, 
analogous to eq.~\loreorder: 
\eqn\bzchib{\eqalign{
&\bar\chi_0(\ahat,M,N)=\ahat\bar\chi_0(M,N)
+\ahat^2\chi_1^{\beta_0}(M,N)+O(\beta_0^2\ahat^2)
\cr 
& \quad\chi_1^{\beta_0}(M,N)=-\beta_0\smallfrac{n_c}{\pi}
\big(\psi^\prime(1-M+\smallfrac{N}{2})
-\smallfrac{1}{(1-M+\smallfrac{N}{2})^2}\big)=-\beta_0
\smallfrac{n_c}{\pi}\psi^\prime(2-M+\smallfrac{N}{2}),\cr
}}
where $\bar\chi_0(M,N)$ is given by eq.~\chicollos.

The second is due to ordering in the collinear and anticollinear
contributions $\chi_s$ in eq.~\simmLORC, which can be determined by
expanding about $M=\half$, just as in  eq.~\rcreshuf, 
but now retaining commutator contributions up to order $\ahat$. Note that,
because the symmetry is broken by
operator ordering, we now also get a cubic contribution. We obtain
\eqn\rcreshufbz{\chi_s\big(\ahat(M+\smallfrac{N}{2})^{-1}\big)+ 
\chi_s\big((1-M+\smallfrac{N}{2})^{-1}\big)\ahat
=\bar\chi^q_{s}(\ahat,M,N)+
\chi_{s}^{\beta_0}(\ahat,M,N)+O\big((\ahat\beta_0)^2\big)}
where $\bar\chi^q_{s}(\ahat,M,N)$ is given by eq.~\chisqdef\ and the
commutator correction is
\eqn\chibs{\chi_{s}^{\beta_0}(\ahat,M,N)=  \bar c^{\beta_0}(\ahat,N) +
\bar \delta^{\beta_0}(\ahat,N) 
\left(M-\half \right)
+\half \bar
\kappa^{\beta_0}(\ahat,N) \left(M-\half \right)^2
+O\big((M-\half)^3\big),}
where
\eqn\rccoeff{\eqalign{
\bar c^{\beta_0}(\ahat,N)&=-2\beta_0\frac{\ahat^2}{1+N}
c^\prime\left(\frac{\ahat}{1+N}\right), 
\cr
\bar \delta^{\beta_0}(\ahat,N)&
=-\beta_0\frac{\ahat^2}{(1+N)^2}\kappa'\left(\frac{\ahat}{1+N}\right),
\cr
\bar \kappa^{\beta_0}(\ahat,N)&=-2\beta_0\frac{\ahat^2}{(1+N)^3}
\left[8c'\left(\frac{\ahat}{1+N}\right) 
- 4 \delta'\left(\frac{\ahat}{1+N}\right)
+ \kappa'\left(\frac{\ahat}{1+N}\right)
\right], \cr
}}
where the functions $c(\as)$, $\delta(\as)$ and $\kappa(\as)$ are
defined in eq.~\chisexp, and the prime denotes differentiation with respect
to their argument.

In summary, the quadratic approximation to the operator  $\bar\chi_{\Sigma\,LO}(\ahat,M,N)$ 
eq.~\simmLORC\ is equal to that which is found when $\as$ is an
ordinary commuting function, plus the two commutator corrections
$\chi^{\beta_0}_1$ eq.~\bzchib\ and $\chi^{\beta_0}_s$ eq.~\chibs, up
to $O((\ahat\beta_0)^2)$ corrections.
Note that because of the cubic term,  the minimum  
is shifted away from
$M=\half$. For instance, the minimum of the symmetrized collinear
contributions is shifted to $M_0$ given by
\eqn\minshift{ M_0=\half-
\frac{\bar\delta^\beta_0(\as,N)}{\bar\kappa(\as,N)
+\bar\kappa^{\beta_0}(\as,N)}.}
Re-expanding the symmetrized collinear contribution about $M_0$ one gets
\eqn\chishift{\eqalign{&
\bar\chi^q_{s}(\ahat,M,N)+
\chi_{s}^{\beta_0}(M,N)= \bar c(\ahat,N)+\bar
c^{\beta_0}(\ahat,N)- 
\half\frac{\bar\delta^\beta_0(\as,N)^2}{\bar\kappa(\as,N)
+\bar\kappa^{\beta_0}(\as,N)}\cr&\qquad\qquad\qquad
+\half (\bar \kappa(\ahat,N)+\bar\kappa^{\beta_0}(\as,N)) \left(M-M_0
\right)^2+O\big(\left(M-M_0\right)^3\big) 
}}

Because operator ordering is immaterial in subleading contributions,
the quadratic approximation at NLO can be obtained by simply adding
the leading-order
commutator corrections $\chi^{\beta_0}_1$ eq.~\bzchib\ 
and $\chi^{\beta_0}_s$ eq.~\chibs~
to the NLO kernel $\bar\chi_{\sigma}(\ahat,M,N)$ eq.~\kisym. The 
quadratic kernel can then be
determined in two equivalent ways. One possibility is to expand 
$\bar\chi_{\sigma}(\ahat,M,N)$ about $M=\half$, and determine the shifted
minimum due to $\beta_0$ corrections and the expansion about it using
eq.~\chishift, with $\bar c$ and $\bar\kappa$ replaced by the
intercept and curvature of the quadratic expansion of
$\bar\chi_{\sigma}(\ahat,M,N)$ about $M=\half$, and $\bar
c^{\beta_0}$, $\bar\delta^{\beta_0}$ and $\bar\kappa^{\beta_0}$
replaced by the corresponding parameters of the quadratic expansion of
$\chi^{\beta_0}_1+\chi^{\beta_0}_s$. This then gives a result of the
form of eq.~\chishift, which can be put on shell by letting
$N=\chi_{\sigma}(\as,M)$, in turn obtained putting on shell the kernel
$\chi_{\sigma}(\as,M,N)$ eq.~\kiasym. A simpler option for the sake of
numerical applications (which we will adopt for the computation of
resummed anomalous dimensions and splitting functions) is to just 
put on shell the kernel
\eqn\chibzos{\chi_{\sigma}(\as,M,N)+\ahat^2\chi^{\beta_0}_1(M,N)
+\chi^{\beta_0}_s(M,N),}
 and then
determine numerically the location of its minimum $M_0$ and the
quadratic approximation about it. Of course, the two procedures are
equivalent up to subleading corrections.

All the discussion so far applies to the unintegrated distribution,
which evolves with a symmetric kernel (provided the symmetric operator
ordering on the left-hand side of eq.~\rcreshuf\ is adopted). It is
easy to switch to the kernel for the unintegrated distribution by
using eq.~\commiuis\ in reverse. Specifically, given the quadratic
approximation 
\eqn\chisonsh{\bar\chi^q_{\sigma}(\ahat,M)
= \bar c_{\sigma}(\ahat,N) 
+\half \bar\kappa_{\sigma}(\ahat,N) \left(M-\half
\right)^2}
to the kernel obtained putting on--shell the symmetric--variable
kernel $\bar\chi_{\sigma}(\ahat,M,N)$ eq.~\kisym, switching to the
integrated distribution  
generates a further
$O(\ahat^2 \beta_0)$ correction to the kernel
of the form
\eqn\quadiui{\chi^{\beta_0}_i=c^{\beta_0}_i(\ahat,N,M) + \half
\kappa^{\beta_0}_i(\ahat,N,M) \left(M-\half\right)^2,}
with
\eqn\quadiuic{\eqalign{
 c^{\beta_0}_i(\ahat,N,M)&=\left(M+\smallfrac{N}{2}\right)^{-1}\left[
 c_\sigma(\ahat,N),M \right]\cr
&=\ahat^2\beta_0\left(M+\smallfrac{N}{2}\right)^{-1}
 \frac{\partial c_\sigma(\ahat,N)}{\partial \ahat},\cr
 \kappa^{\beta_0}_i(\ahat,N,M)&=\left(M+\smallfrac{N}{2}\right)^{-1}\left[\kappa_\sigma(\ahat,N),M \right]\cr
&=\ahat^2\beta_0\left(M+\smallfrac{N}{2}\right)^{-1}
 \frac{\partial \bar\kappa_\sigma(\ahat,N)}{\partial \ahat},
\cr
}}
where we have taken into account the fact
that eq.~\commiuis\ is
 expressed in asymmetric variables so a further shift is necessary to
 bring it to symmetric variables as in eq.~\simmLORC. Because the
 coefficients $c^{\beta_0}_i$ and $\kappa^{\beta_0}_i$ are
 $M$--dependent, they must be expanded about $M=\half$. Using this
 expansion in eq.~\quadiui\ and keeping terms up to second order, leads
 to a set of contributions to the intercept, linear term and curvature
 which  must be added to those of eq.~\rccoeff, and treated in
 the same way in order to obtain the quadratic approximation to the
 kernel for the integrated distribution. 

Once the coefficients of the quadratic expansion about $M_0$ have been
determined, they can be used to determine the Bateman anomalous
dimension eq.~\losadad, as well as its asymptotic expansion in the
$\beta_0\to0$ limit eqs~\ssbat\ and \gambzdef.
We thus obtain the full NLO resummed anomalous dimension, as discussed
in ref.~\refs{\sxrun,\runph} and summarized in sect.~2, by 
determining the perturbative running-coupling dual anomalous dimension
$\gamma_{\sigma\,NLO}$, and then combining it with the Bateman
resummation determined in the quadratic approximation, and 
subtracting the double counting.

The perturbative running coupling dual of $\chi_{\Sigma}(\as,M,N)$
eq.~\kisym\ is constructed as follows.
 We start from the observation
that~\sxrun\ the dual of the NLO double--leading $\chi$  coincides
with the NLO double--leading $\gamma$, up to subleading
terms. However,  recall that in order to
determine the anomalous dimension we must first switch back to the
integrated gluon distribution. Upon this transformation, $\chi_s$
changes according to eq.~\commiuis, while for $\chi_0$ we have
\eqn\commiuiz{\eqalign{
\ahat \chi_0(M) {\cal
G}(\xi,M)&= \ahat \chi_0(M) M  G(\xi,M)\cr 
&=M \left(\ahat\chi_0(M)+\ahat^2
\beta_0\frac{\chi_0(M)}{M}+O(\ahat^3)\right) G(\xi,M).}}
Now, we note
that the dual anomalous dimension to the kernel 
$\chi_s(\ahat M^{-1})+\ahat \chi_{ss}(\ahat M^{-1})$ which we get
after switching back to the integrated level is just, by
construction, the NLO anomalous dimension as given by eq.~\gammadef\
up to order $\as^2$. On the other hand, the dual of the $O(\ahat^2)$ kernel 
eq.~\flresult\
is given by eq.~\rccor, supplemented by the $O(\ahat^2)$ term of eq.~\commiuiz.
 Note
that the result holds with the operator ordering of
eq.~\flresult.
The $O(\ahat)$
contribution to $\chi_\sigma$ can be brought to this form by
using eq.~\flresultreord, and subtracting the double pole from the
$\beta_0$ term of eq.~\flresultreord, which is a double--counting
contribution
between $\chi_1$ and $\chi_s((1-M)^{-1}\ahat)$. In practice, this means
that the $\beta_0$ term of eq.~\flresultreord\ should be replaced by
$\chi_1^{\beta_0}$ eq.~\bzchib.

 Putting everything together, the perturbative running
coupling dual to $\chi_{\sigma\,NLO}(M)$ differs from the
fixed-coupling dual in three respects: first, the $O(\ahat)$
contribution to $\chi_\Sigma$ must be reordered so that $\ahat$ is to
the left, and second it must be supplemented by the contribution
eq.~\commiuiz\ which switches to the integrated distribution; finally, 
the fixed coupling duality must be corrected by the
$O(\beta_0)$ term of eq.~\rccor. We thus get
\eqn\prcdual{\gamma^{rc,\,pert}_{\Sigma\,NLO}(\as(t),N)=
\gamma_{\Sigma\,NLO}(\as(t),N)
-\beta_0\as(t)\left[\frac{\chi_0^{\prime\prime}(\gamma_s(\as/N))
\chi_0(\gamma_s(\as/N))}{2[\chi_0^\prime(\gamma_s(\as/N))]^2}-1\right],} 
where the subtraction in the term in square brackets subtracts the
double-counting between this term and $\gamma_0$, i.e. it is part of
the standard subtraction $e_0$ in eq.~(22) of ref.~\sxres.
The anomalous dimension $\gamma_{\Sigma\,NLO}(N,\as(t))$ in eq.~\prcdual\
is the naive (fixed coupling)
dual eq.~\dualdef\ 
of $\chi_{\Sigma\,NLO}$ obtained by putting on shell the kernel in
asymmetric variables,
\eqn\allchi{
\bar\chi_{\Sigma\,NLO}(M-\smallfrac{N}{2},N)+\ahat^2\chi_1^{\beta_0}(M-\smallfrac{N}{2},N)+\ahat^2\beta_0\left(\frac{\chi_0(M)}{M}-\frac{n_c}{\pi
M^2}\right),} 
where
$\bar\chi_{\Sigma\,NLO}(M,N)$ is given by eq.~\kisym\  as discussed in
this section, 
$\chi_1^{\beta_0}(M)$ is given by eq.~\bzchib, and the  term in
brackets is the transformation to the kernel for the integrated
distribution, with the double counting between it and
$\chi_s(M^{-1}\ahat)$ subtracted.

We finally obtain the full resummed anomalous dimension by combining
this result with the Bateman running coupling resummation:
\eqn\rcnlores{\eqalign{&\gamma^{rc}_{\Sigma\,NLO}(\as(t),N)
=\gamma^{rc,\,pert}_{\Sigma\,NLO}(\as(t),N)+
\gamma^B(\as(t),N)- 
\gamma^B_s(\as(t),N)\cr&\qquad- 
\gamma^B_{ss}(\as(t),N)-\gamma^B_{ss,0}(\as(t),N)+\gamma_{\rm match}(\as(t),N)
+\gamma_{\rm mom}(\as(t),N).}}
In this equation, the Bateman anomalous dimension is given by
eq.~\batandim, computed with the curvature $c$ and intercept $\kappa$
determined 
from the quadratic expansion of the on-shell form of the kernel eq.~\chibzos\
discussed above. However, in oder to ensure complete cancellation of the
singularities, the asymptotic double counting terms
$\gamma^B_s(\as(t),N)$, $\gamma^B_{ss}(\as(t),N)$ and 
$\gamma^B_{ss,0}(\as(t),N)$   should be determined 
with values of the parameters $c$ and $\kappa$ which differ from those
used in $\gamma^B$ itself by
sub-subleading terms. Specifically, $\gamma^B_s(\as(t),N)$  should be 
computed with
the values of  $c$ and $\kappa$ which characterize $\gamma_{\Sigma
NLO}$ eq.~\prcdual, namely, the intercept and curvature of the 
quadratic expansion
of the kernel $\chi(\gamma_{\Sigma
NLO})$ related to $\gamma_{\Sigma
NLO}$ by fixed-coupling duality eq.~\dualdef. These differ from the 
values of $c^{\beta_0}$ of $\kappa^{\beta_0}$ of the quadratic 
expansion because of the
contribution to the latter from $\chi_s^{\beta_0}$ eq.~\chibs. 
In practice, we eliminate this 
subsubleading singularity mismatch through the term 
\eqn\sqrtmatch{\eqalign{\gamma_{\rm match}(\as(t),N)&=\sqrt{\frac{N-c}{\half\kappa}}
               -\sqrt{\frac{N-c^{\beta_0}}{\half\kappa^{\beta_0}}}\cr
&\qquad               -\sqrt{\frac{N+1}{\half\kappa}}
               +\sqrt{\frac{N+1}{\half\kappa^{\beta_0}}}
              +\frac{1+c}{\sqrt{2\kappa(N+1)}}
             -\frac{1+c^{\beta_0}}{\sqrt{2\kappa^{\beta_0}(N+1)}}:}}
the first two terms shift the square root cut, and the 
remaining ones ensure that at large $N$ the resummed
anomalous dimension eq.\rcnlores\ 
reproduces the NLO GLAP result.
Furthermore, $\gamma^B_{ss}(\as(t),N)$ is computed with the intercept and
curvature of the leading order BFKL kernel $\chi_0$ eq.~\kizero, which
ensures cancellation of the pole at $M=\half$ of the $O(\beta_0)$ term
in eq.~\prcdual\ which motivates the running coupling resummation, as
discussed in refs.~\refs{\sxrun,\runph}\ and summarized in
sect.~2. The values used in $\gamma^B$ differ from these due to the
inclusion of all the subleading corrections discussed above, but this
is immaterial because $\gamma^B_{ss}$ is already subleading.
Finally, the term $\gamma_{\rm mom}$ serves
the purpose of improving the matching to large $N$, by removing terms
which survive at the momentum-conservation point $N=1$.
These terms are of course sub-subleading, so they
could be in principle omitted. In practice $\gamma_{\rm
mom}$ is constructed as in eq.~\momsubnew, with $d_m$ readjusted to
ensure that $\gamma^{rc}_{\Sigma\,NLO}(1)=0$.

\newsec{Results for the Anomalous Dimension and the Splitting Function}

We are now finally in a position to collect and discuss our
results. Our goal was to derive for the singlet anomalous dimension
and for the associated splitting function expressions that at large
$N$ or large $x$ reduce to the familiar perturbative GLAP results,
and at small $N$ or small $x$ are improved by including in a
suitably resummed form the information from the BFKL kernels with a
procedure that takes into full account the constraints from momentum
conservation and symmetrisation and includes running coupling
effects. As mentioned in the introduction, all results presented in
this section are computed with $n_f=0$ in order to minimize the impact
of issues of scheme dependence.

\topinsert
\vbox{
\epsfxsize=12truecm
\centerline{\epsfbox{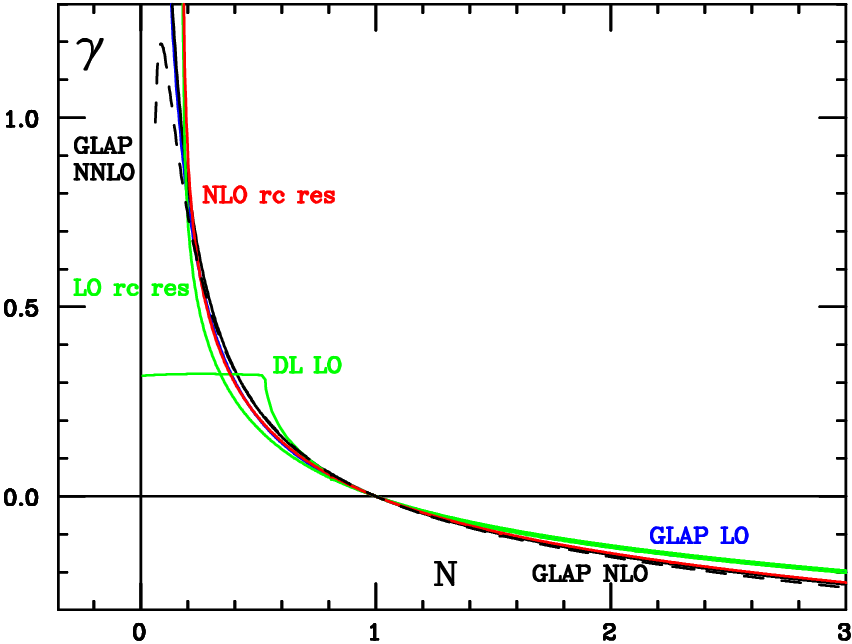}}
\bigskip
\hbox{
\vbox{\footnotefont\baselineskip6pt\narrower\noindent Figure 7: Our
most accurate prediction, denoted `NLO rc res', eq.~\rcnlores, is
compared  to the LO, NLO and NNLO GLAP perturbative results and to the 
DL LO approximation, eq.~\dlgam. }}\hskip1truecm}
\endinsert 

\topinsert
\vbox{
\epsfxsize=12truecm
\centerline{\epsfbox{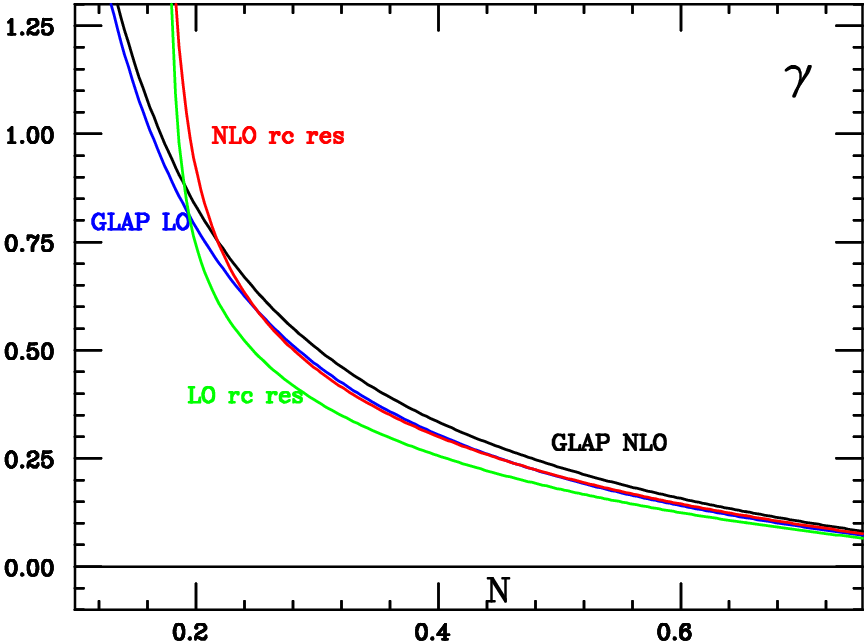}}
\bigskip
\hbox{
\vbox{\footnotefont\baselineskip6pt\narrower\noindent Figure 8: The
NLO resummed prediction eq.~\rcnlores, denoted `NLO rc res' is
compared to the LO and NLO GLAP perturbative results and to
the leading 
approximation, eq.~\rcres, denoted by `LO rc res'. }}\hskip1truecm}
\endinsert 

The results for the anomalous dimension are presented in figs.~7-8. In
fig.~7 our most accurate prediction, denoted `NLO rc res', is compared
with the LO, NLO and NNLO GLAP perturbative results and with the DL LO
approximation. The curve labelled  `NLO rc res' shows the anomalous
dimension, which  includes the information from
the NLO BFKL kernel and perturbative anomalous
dimension, to which it reduces at large $N$. It corresponds to the result given
in eq.~\rcnlores. The DL LO curve, obtained from $\gamma_{DL LO}$
given in eq.~\dlgam, corresponds to a naive application of the BFKL
kernel $\chi_0$ to evaluate the small $N$ behaviour. The step
corresponds to the cut starting at the value of $N$ which corresponds
by duality to the minimum of $\chi_0$. It is quite evident that our
final improved anomalous dimension is remarkably close to the GLAP NLO
result down to very small values of $N$, where finally we can see a
departure due to the position of the Bateman pole which is at a small
positive value of $N$ and not at $N=0$ as is the case for GLAP. The
instability of the NNLO GLAP result is due to an unresummed
$\frac{\as^3}{N^2}$ pole, and it is removed by the resummation.

\topinsert
\vbox{
\epsfxsize=12truecm
\centerline{\epsfbox{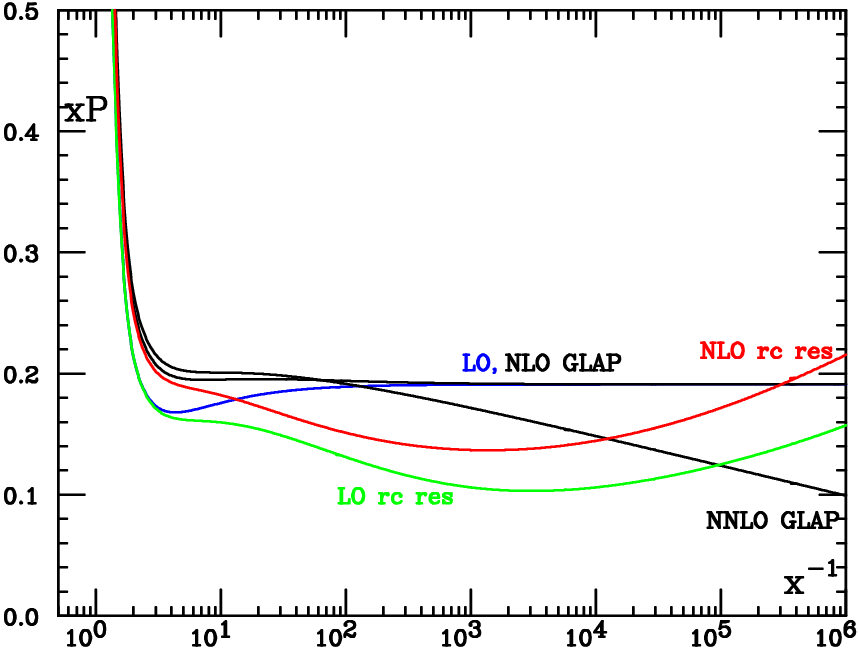}}
\bigskip
\hbox{
\vbox{\footnotefont\baselineskip6pt\narrower\noindent Figure 9: The
LO and NLO resummed splitting function obtained by Mellin
transformation of the anomalous dimensions eq.~\rcres\ and
eq.~\rcnlores\ displayed in
fig.~8 (labelled `rc res') are compared to the LO, NLO and NNLO GLAP perturbative
splitting functions (note that the plot shows $xP$).  }}\hskip1truecm}
\endinsert 
Figure~8 shows a magnification of the small $N$ region, and also
compares the NLO resummed result to its LO counterpart eq.~\rcres.
The stability of the resummed perturbative expansion is manifest in
the small difference between the LO and NLO results, which however are
in perfect agreement with the respective GLAP curves for large $N\gsim
1$. 
At medium-small $N$ the NLO resummed curve
is in fact rather closer to the unresummed one than the LO resummed.
At very small $N$ the LO and NLO resummed curves are almost
indistinguishable because of the closeness of the respective Bateman
poles, displayed in Fig.~6, and rise somewhat more steeply than the
unresummed result.

We now consider the splitting function obtained by inverse Mellin
transformation from these anomalous dimension. In fig.~9 the resummed splitting function
curves (labelled `LO rc res' and `NLO rc res') which are our main predictions at
the leading and next-to-leading level, corresponding to the anomalous
dimensions with the same labels in fig.~8, are compared with the LO,
NLO and NNLO GLAP perturbative splitting functions (note that what is
actually plotted is $xP$). The first important feature is the marked
deviation of the curve NNLO GLAP from the approximate $1/x$ behaviour
at small $x$ which is typical of the LO and NLO GLAP singlet splitting
function. This pronounced deviation makes the need for resummation
even more compelling, in the sense that it clearly shows that the
ordinary perturbative expansion is unstable at small $x$. As the bulk
of HERA data is concentrated at $x>10^{-3}-10^{-4}$ the departure from
the NLO splitting functions used in most fits to the data is  not too
large, but, for the perturbative results, the difference NNLO-NLO is
much larger than NLO-LO in the whole range of small $x$. The resummed
NLO prediction, which constitutes our main result, closely follows the
NLO GLAP curve down to $x$ values as small as $x\sim 10^{-5}$, thus
explaining the success of the NLO perturbative QCD fits of
the HERA data. In comparison the LO resummed prediction shows 
more pronounced deviations from the perturbative splitting
function. It is only at very small $x$, $x<10^{-6}$, that there starts
to be a noticeable difference between the perturbative and the resummed
splitting functions. This late take-over of the BFKL regime, softened
by the resummation procedure, is due to the small residue of the
Bateman pole. 

\topinsert
\vbox{
\epsfxsize=12truecm
\centerline{\epsfbox{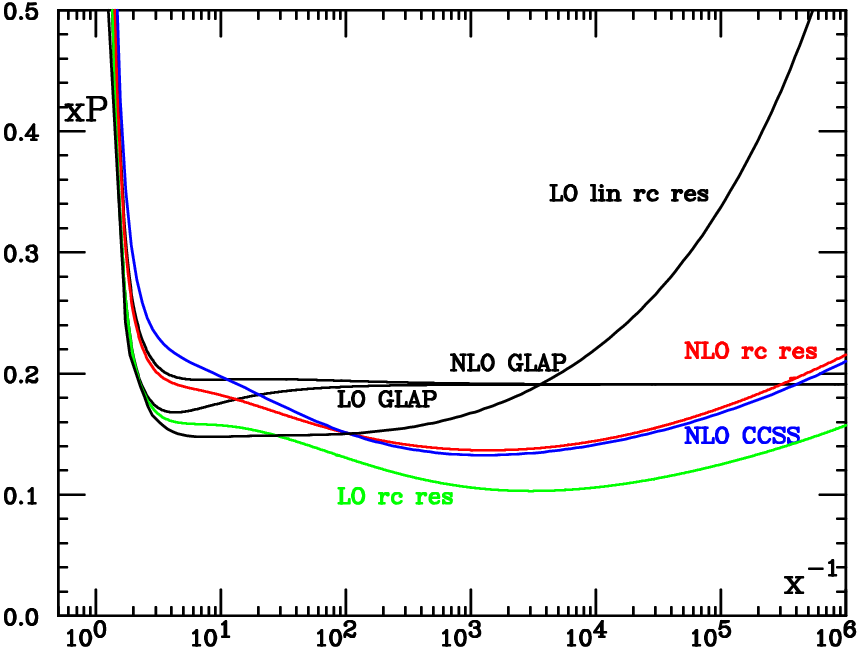}}
\bigskip
\hbox{
\vbox{\footnotefont\baselineskip6pt\narrower\noindent Figure 10: The
LO and NLO resummed splitting function of fig.~9 are compared to
the result obtained in our previous work, ref.~\runph, denoted by `LO lin rc
res' and to the  result of ref.~\ciafresb, labeled  `NLO CCSS' }}\hskip1truecm}
\endinsert  
In fig.~10 we compare our main results for the splitting function,
i.e. the  LO and NLO resummed curves of fig.~9,  to
the result obtained in our previous work, ref.~\runph, denoted by `LO lin rc
res' and with the result (labeled `NLO CCSS') which was obtained
by the authors of ref.\ciafresb\ by a different technique based on the same
physical ingredients. The  prediction of ref.~\runph\ is a
simple analytical result that can be described as the best
approximation that can be derived if only the 1-loop kernels
$\gamma_0$ and $\chi_0$ are known. Note that, since $\chi_0$ is
already symmetric for $M \rightarrow (1-M)$, there is no need of
symmetrisation, while the result shown includes collinear resummation,
momentum conservation and running coupling effects. The rise at small
$x$ is steeper and takes off at larger $x$ than for the more refined
results obtained in this paper and in ref.~\ciafresb, because the symmetrised
resummed DL kernels displayed in figs.~1 and 3 have a smaller value at
the minimum and are much flatter than $\chi_0$. The main advantage of
the new result, however, is its perturbative stability, i.e. the
closeness of LO and NLO results.

The approach of ref.~\ciafresb, although based on similar physical
principles, differs in many respects from ours, most notably in the
fact that the whole approach is based on the BFKL equation, and
attempts to derive the full gluon Green function, instead of focussing
on the determination of anomalous dimensions, as in our approach, which treats
the $\chi$ BFKL kernel and $\gamma$ anomalous dimension on the same
footing. It 
is therefore to some extent unexpected
that our NLO resummed result at small $x$  is very close 
to the NLO CCSS splitting function.  
In fact, the very small difference between our results and
those of CCSS is surely accidental, given the uncertainty involved in
various subleading resummation terms.
Note that even though the approach of ref.~\ciafresb\ does
not yet allow for the inclusion of NLO GLAP terms when $n_f\not=0$,
the CCSS curve of fig.~10, taken from refs.~\heralhcres\ and
computed with $n_f=0$, does include a NLO GLAP contribution, which
when $n_f=0$ vanishes as $x\to0$.
\topinsert
\vbox{
\epsfxsize=12truecm
\centerline{\epsfbox{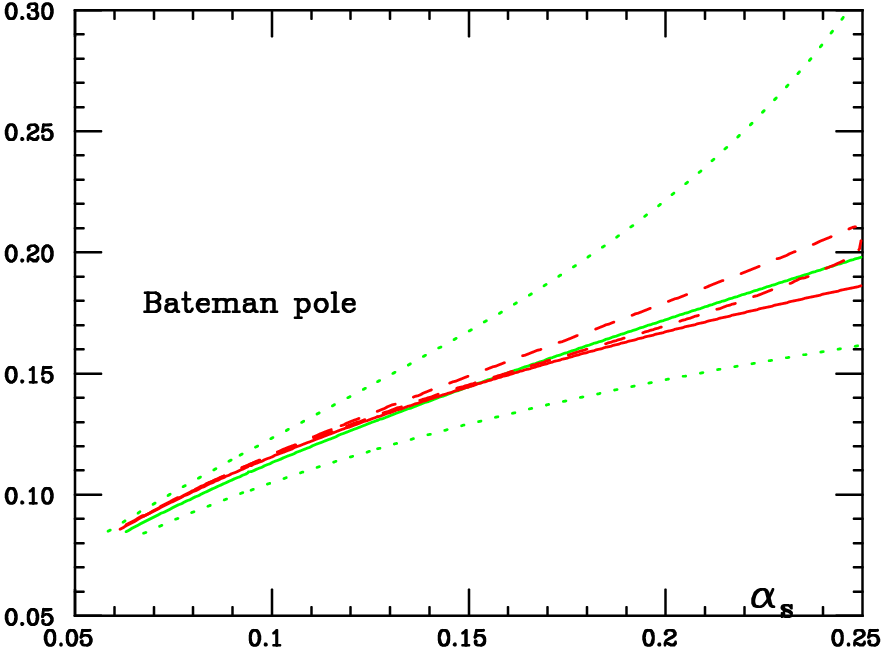}}
\bigskip
\hbox{
\vbox{\footnotefont\baselineskip6pt\narrower\noindent Figure 11: 
Dependence of the rightmost pole $N_B$ of the leading and
next-to-leading order resummed anomalous
dimensions eq.~\rcres,\rcnlores\   on $\alpha_s(k Q^2)$ 
for various values of the renormalization scale $k$. The solid curves
correspond to $k=1$ (lower: LO, higher: NLO), while the dotted (LO)
and dashed  (NLO) pair of curves give the range of variation as
$\smallfrac{1}{4}\le k\le 4$.}}
\hskip1truecm}
\endinsert

A feeling for the uncertainty of our results can be obtained by
determining their dependence on the choice scale $Q^2$, which in turn
may be estimated from the dependence of the position of the rightmost
singularity $N_B$ of the resummed anomalous dimensions eq.~\rcres,\rcnlores\
on the value of $\alpha_s$, which was displayed in fig.~6: the
value of $N_B$ determines the asymptotic small $x$ behaviour 
of the splitting function $x P\tozero {x} x^{-N_B}$. We can also use
this dependence to  estimate the uncertainty on our results due to
subleading higher order terms. To this purpose, we study the
dependence of the value of $N_B(\as)$ on the choice of renormalization
scale. At leading order, we plot $N_B(\as(k Q^2))$, with $N_B$ the
rightmost pole of the leading order $\gamma^{rc}_{\sigma\,LO}$
eq.~\rcres, while at next-to-leading order we plot
\eqn\scalvar{N_B(k,Q^2)=N_B(\as(kQ^2)) + \beta_0 \ln k \,\as^2(Q^2) 
\frac{\partial }{\partial\as} N_B(\as(Q^2)),}
with $N_B$ the rightmost pole of the next-to-leading
order $\gamma^{rc}_{\sigma\,NLO}$
eq.~\rcnlores. In fig.~11 we compare the values of the leading order and
next-to-leading order $N_B$ as a function of $\alpha_s(k Q^2)$ for
$k=1$, $k=4$ and $k=\smallfrac{1}{4}$, while in fig.~12 we show the
dependence of these quantities 
on $k$
when $Q^2$ is chosen so that $\as(Q^2)=0.2$. We see that the
scale dependence of the leading--order result is about as large as the
difference between  leading and next-to-leading order results, whereas
the scale dependence of the next-leading--order result is smaller by
about a factor four. This is somewhat larger than the difference
between our NLO result and that of ref.~\ciafresb, as shown in fig.~10.
Also, it appears from fig.~12 that whereas at leading order
$N_B$ is monotonic as a function of the scale parameter $k$, at
next-to-leading order the dependence of $N_B$ on $k$ is stationary
around $k\sim 1$, confirming that indeed 
$Q^2$ is the appropriate scale for this process. We conclude from the 
significantly reduced dependence on renormalization scale at NLO that 
we have indeed succeeded in constructing a stable resummed perturbative 
expansion in the small $x$ region. If we were to go to NNLO in this 
expansion, we would expect the scale dependence to be reduced still 
further.
\topinsert
\vbox{
\epsfxsize=12truecm
\centerline{\epsfbox{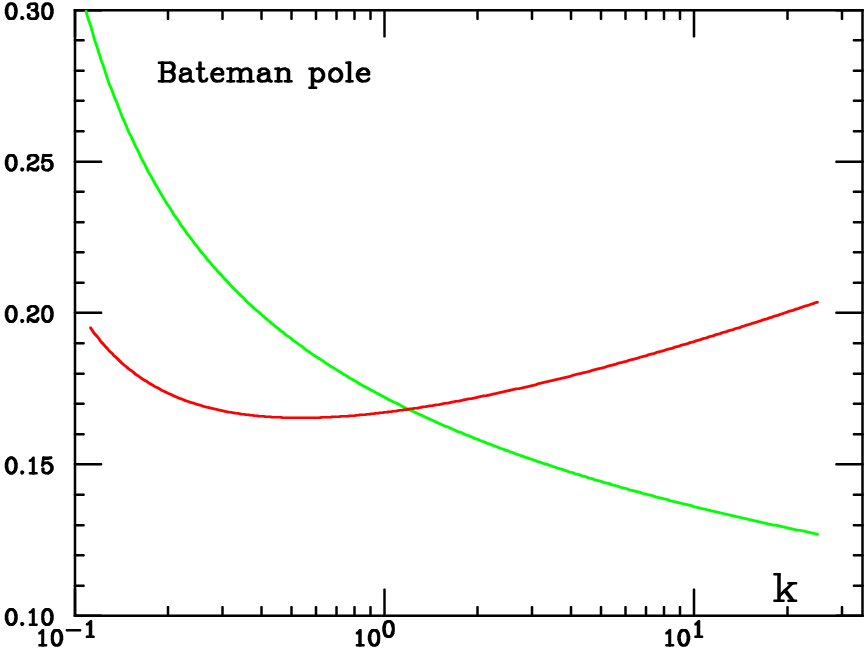}}
\bigskip
\hbox{
\vbox{\footnotefont\baselineskip6pt\narrower\noindent Figure 12: 
Dependence of the rightmost pole $N_B$ of the leading and
next-to-leading order resummed anomalous
dimensions eq.~\rcres,\rcnlores\ 
on the renormalization scale
$k$ when $\as(Q^2)=0.2$ (lower at large $k$: LO, higher: NLO).
}}\hskip1truecm} 
\endinsert

\newsec{Conclusion}

In this paper we have presented in detail a complete procedure to
construct an improved singlet anomalous dimension or splitting
function that reduces at large $x$ to the NLO perturbative
approximation but also contains all resummed small $x$ improvements
from NLO BFKL kernels. The problem of explaining the apparent
smallness of the small $x$ corrections in the HERA data region in
spite of the wild behaviour of the BFKL perturbative expansion is
solved. The main physical reason for the moderate impact of the small
$x$ corrections is the duality relation that implies that in the large
domain where both $Q^2$ and $1/x$ are large the GLAP and the BFKL
expansions are both valid and know about each other. Thus the corresponding
leading twist terms must match and indeed a better convergence is
obtained by the `double leading (DL)' expansion where at each level
the BFKL (GLAP) perturbative terms are used to resum the corrections
to the GLAP (BFKL) kernel in a symmetric way. The constraints from
momentum conservation and from the underlying symmetry of the
gluon-gluon BFKL scattering kernel under interchange of the two gluons
are used to construct a stable and physically motivated expansion at
each level. An important effect is also obtained from the running
coupling corrections which are large and considerably soften the small
$x$ asymptotic behaviour. The technical implementation of the running
coupling corrections in our approach is realized via a quadratic
approximation of the BFKL kernel in the DL approximation near its
central minimum followed by the analytic solution of the corresponding
differential equation in terms of a Bateman function. The systematic
use of duality and the Bateman method are the main qualitative
differences with respect to the approach of ref.\ciafresb\ which uses a
different path to reach similar conclusions from the same basic
physical principles. 

The main results of our work for the singlet anomalous dimension and
the corresponding splitting function are displayed in figs. 7-11 where
the comparison with the results of ref.\ciafresb\ is also shown. In the region
of the bulk of HERA data for $x > 10^{-4}$ the corrected splitting
function at NLO level shows a moderate depletion with respect to the
perturbative NLO approximation. At still smaller values of $x$ 
the asymptotic regime eventually takes over and a powerlike increase 
of the splitting
function is observed, though at a much lower rate than that naively
expectated from the value of the Lipatov exponent.  

We recall once again that to make direct connection to the structure functions
we need to include the coefficient functions. This in principle is 
relatively easy, because the coefficient
functions are available, even though properly accounting for scheme
choices,  diagonalization of the anomalous dimension  matrix, and
matching to large $x$ coefficient functions is a laborious task, as
discussed in ref.~\sxphen. The effects of the coefficient functions
are subdominant in the factorization scheme adopted here, but they
deserve a full investigation.

In conclusion we think that by now the problem of the small $x$
behaviour of the singlet splitting functions has been well understood
in its physical principles and that the technical tools have been
satisfactorily developed. The next step is to make a direct comparison
of the resulting theoretical predictions with the data.

{\bf Acknowledgements:} We thank M.~Ciafaloni and especially G.~Salam
for innumerable discussions and exchanges. S.F. thanks E.~Iancu,
D.~Triantafyllopoulos  and B.~Pire for discussions and
the Centre de Physique Th\'eorique of Ecole Polytechnique, where part
of this work was done, for hospitality. This work was funded in part 
by PPARC and the Scottish Universities Physics Alliance.

\appendix{A}{}

To evaluate $\chi_1(M)$ we use the result 
\eqn\id{\eqalign{\varphi(M)&\equiv -\int_0^1 \frac{dx}{1+x}(x^{M-1}+x^{-M})
\int_x^1 \frac{dt}{t}\ln (1-t)\cr
&= \frac{\pi^3}{6\sin\pi M} - \frac{\pi^2}{6M(1-M)}\cr
&\qquad-\frac{1}{M^2}(\psi(1)-\psi(1+M))-\frac{1}{(1-M)^2}(\psi(1)-\psi(2-M))
+\varphi_s(M),\cr}}
where
\eqn\phiess{\varphi_s(M)=\sum_{r=1}^\infty c_r (M-\half)^{2(r-1)},}
and
\eqn\coeffs{c_r = -2\sum_{k=1}^\infty \sum_{m=1}^\infty 
\frac{(-)^m}{k^2}\frac{1}{(m+k+\half)^{2r-1}},}
so the sum over $r$ has infinite radius of convergence. If we further note 
that
\eqn\idone{\frac{2\pi}{\sin\pi M} = \psi(\half+\smallfrac{M}{2})
-\psi(\smallfrac{M}{2}) + (M \leftrightarrow 1-M),}
\eqn\idtwo{\frac{4\pi^2 \cos\pi M}{\sin^2\pi M} = \psi'(\smallfrac{M}{2})
-\psi'(\half+\smallfrac{M}{2}) - (M \leftrightarrow 1-M),}
then we may then use eq.~(14) of ref.~\fl\ to write $\chi_1(M)
=\smallfrac{n_c^2}{4\pi^2}\delta(M)-\beta_0\smallfrac{n_c}{\pi}\psi'(M)$ as 
\eqn\chionez{\eqalign{\chi_1(M) &= -\half\beta_0\smallfrac{n_c}{\pi}
(\smallfrac{\pi^2}{n_c^2}\chi_0(M)^2 -\psi'(M)-\psi'(1-M))\cr
&\qquad +\smallfrac{n_c^2}{4\pi^2}\Big[(\smallfrac{67}{9}-\smallfrac{\pi^2}{3}-
\smallfrac{10n_f}{9n_c})(\psi(1)-\psi(M))+ 3\zeta (3) +\psi''(M)\cr
&\qquad + 4(\pi^2(\psi(\half+\smallfrac{M}{2})-\psi(\smallfrac{M}{2}))
+\frac{6\pi^2}{M}+\frac{1}{M^2}(\psi(1)-\psi(1+M))-\half \varphi_s(M))\cr
&\qquad + \frac{1}{4(1-2M)} \Big(3+(1+\smallfrac{n_f}{n_c^3})
\frac{(2+3M(1-M))}{(3-2M)(1+2M)}\Big)
(\psi'(\half+\smallfrac{M}{2})-\psi'(\smallfrac{M}{2}))\cr 
&\qquad\qquad +(M \leftrightarrow 1-M)\Big].\cr
}}
The advantage of using this expression is that all terms in square
brackets  
are regular for $\Re M > \half$ (note that the poles at $M=\half$, 
$M=\smallfrac{3}{2}$ and $M=-\half$ in the last term are all removable: 
they cancel against similar poles in the $M\to 1-M$ piece). 
Thus we may extend them off-shell 
following the same procedure as at LO. The first term may be treated by 
the simple observation that up to subleading terms we may substitute 
$\as\chi_0(M)=N$, which is trivially extended off-shell as 
$N=\as\bar\chi_0(M,N)$. The final result for $\bar\chi_1(M,N)$ is then 
(in asymmetric variables for simplicity) 
\eqn\chioneoff{\eqalign{\bar\chi_1(M-\smallfrac{N}{2},N)  
&= -\half\beta_0 \smallfrac{n_c}{\pi}
(\smallfrac{\pi^2}{n_c^2}\bar\chi_0(M-\smallfrac{N}{2},N)^2
 -\psi'(M)-\psi'(1-M+N))\cr
&\qquad +\smallfrac{n_c^2}{4\pi^2}\Big[(\smallfrac{67}{9}-\smallfrac{\pi^2}{3}-
\smallfrac{10n_f}{9n_c})\big\{\half(\psi(1)+\psi(1+N))-\psi(M)\big\}
+ 3\zeta (3) +\psi''(M)\cr
&\qquad + 4\big\{\pi^2(\psi(\half+\smallfrac{M}{2})-\psi(\smallfrac{M}{2}))
+\smallfrac{6\pi^2}{M}+\smallfrac{1}{M^2}(\psi(1)-\psi(1+M))
-\half \varphi_s(M)\big\}\cr
&\qquad + \smallfrac{3}{4(1-2M)} 
(\psi'(\half+\smallfrac{M}{2})-\psi'(\smallfrac{M}{2})
+\psi'(\quarter)-\psi'(\smallfrac{3}{4}))\cr 
&+  \smallfrac{1}{16}(1+\smallfrac{n_f}{n_c^3})
(2+3M(1-M))\Big\{ \smallfrac{1}{(1-2M)}
(\psi'(\half+\smallfrac{M}{2})-\psi'(\smallfrac{M}{2})
+\psi'(\quarter)-\psi'(\smallfrac{3}{4}))\cr 
&\qquad + \smallfrac{1}{2(1+2M)}
(\psi'(\half+\smallfrac{M}{2})-\psi'(\smallfrac{M}{2})
+\psi'(-\quarter)-\psi'(\smallfrac{1}{4}))\cr 
&\qquad + \smallfrac{1}{2(3-2M)}
(\psi'(\half+\smallfrac{M}{2})-\psi'(\smallfrac{M}{2})
+\psi'(\smallfrac{3}{4})-\psi'(\smallfrac{5}{4}))\Big\}\cr 
&\qquad + \smallfrac{3}{512}(1 +\smallfrac{n_f}{n_c^3})(\psi'(-\quarter)
-\psi'(\smallfrac{5}{4})+3\psi'(\smallfrac{3}{4})-3\psi'(\quarter))N\cr
&\qquad\qquad +(M \leftrightarrow 1-M+N)
\Big].\cr
}}
where in the term in curly brackets we have taken care to ensure that we 
do not spoil the cancellation of the poles at $M=\pm\half$, 
$M=\smallfrac{3}{2}$ by adding in subleading 
(ie $O(N)$) terms: the growth of these extra terms at large $N$ must then 
be subtracted off again, and this is done in the last line.
The result in symmetric variables may be found by the simple change of 
variables $M\to M+\smallfrac{N}{2}$.

The off-shell extension of eq.~\chitil\ is then straightforward:
\eqn\chitilos{\breve\chi_1(M,N)
=\bar\chi_1(M,N)
-\half\bar\chi_0(M,N)\smallfrac{n_c}{\pi}
\left[2\psi'(1)-\psi^\prime(M+\smallfrac{N}{2})-
\psi^\prime(1-M+\smallfrac{N}{2})\right].}
The collinear subtraction eqn\chibaronedef\ is then
\eqn\chibaronedefos{\tilde \chi_1(M,N)
= \breve\chi_1(M,N)-g_1\Big(\smallfrac{1}{M+\smallfrac{N}{2}}+
\smallfrac{1}{1-M+\smallfrac{N}{2}}\Big)
-g_2\Big(\smallfrac{1}{\big(M+\smallfrac{N}{2}\big)^2}+
\smallfrac{1}{\big(1-M+\smallfrac{N}{2}\big)^2}\Big),}
with coefficients $g_i$ as before, eq.~\gcoeff, which render 
in particular $\tilde \chi_1(M-\smallfrac{N}{2},N)$ finite in the 
collinear limit $M\to 0$. To improve the matching to GLAP at large $N$, we
may further add to $\tilde \chi_1(M-\smallfrac{N}{2},N)$ a 
manifestly subleading contribution $-\tilde \chi_1(-\smallfrac{N}{2},N)
+\tilde \chi_1(0,0)$ to remove spurious constant terms at large $N$, and thus 
spurious $O(1/N)$ terms in the anomalous dimension.

\footatend\vfill\supereject\immediate\closeout\rfile\writestoppt
\baselineskip=14pt\centerline{{\bf References}}\bigskip{\frenchspacing%
\parindent=20pt\escapechar=` \input refs.tmp\vfill\eject}\nonfrenchspacing
\vfill\eject
\bye

\bigskip
\footatend\vfill\supereject\immediate\closeout\rfile\writestoppt
\baselineskip=14pt\centerline{{\bf References}}\bigskip{\frenchspacing%
\parindent=20pt\escapechar=` \input refs.tmp\vfill\eject}\nonfrenchspacing
\vfill\eject
\bye